\documentclass[useAMS,usenatbib,letterpaper]{mn2e}
\usepackage{graphicx}
\usepackage{epsfig}
\usepackage{natbib}
\usepackage{lscape}
\usepackage{subfigure}
\usepackage{longtable}
\usepackage{verbatim}
\usepackage{amssymb,amsmath}
\DeclareMathVersion{bold}

\newcommand{\enorm}{E/\langle E\rangle}
\newcommand{\se}{\langle s_E\rangle}
\newcommand{\bi}[1]{\emph{\textbf{#1}}}
\newcommand{\eon}{E_{\rm on}}
\newcommand{\eoff}{E_{\rm off}}
\newcommand{\pln}{P_{\ell}}
\newcommand{\pg}{P_{\rm g}}
\newcommand{\sln}{\sigma_{\ell}}
\newcommand{\sg}{\sigma_{\rm g}}
\newcommand{\mln}{\mu_{\ell}}
\newcommand{\mg}{\mu_{\rm g}}
\newcommand{\rj}{R_{\rm j}}
\newcommand{\mj}{m_{\rm j}}

\title[Single-pulse properties of 315 pulsars]{The High Time Resolution Universe Survey -- V: Single- pulse energetics and modulation properties of 315 pulsars}

\author[S. Burke-Spolaor et al.]{S. Burke-Spolaor$^{1,2}$\thanks{Email: sarah.burke-spolaor@jpl.nasa.gov},
S. Johnston$^{1}$,
M. Bailes$^{3,4}$,
S.~D. Bates$^{5,9}$,
N.~D.~R. Bhat$^{3,4}$,
\newauthor
M. Burgay$^{6}$,
D.~J. Champion$^{7}$,
N.~D'Amico$^{6}$,
M.~J. Keith$^{1}$,
M. Kramer$^{7}$,
L. Levin$^3$,
\newauthor
S. Milia$^{6,8}$,
A. Possenti$^{6}$,
B. Stappers$^{9}$,
W. van Straten$^{3,4}$\\
$^1$Australia Telescope National Facility, CSIRO, P.O. Box 76, Epping, NSW 1710, Australia\\
$^2$NASA Jet Propulsion Laboratory, M/S 138-307, Pasadena CA 91106, USA\\
$^3$Swinburne University of Technology, Centre for Astrophysics and Supercomputing Mail H30, P.O Box 218, VIC 3122, Australia\\
$^4$ARC Centre for All-Sky Astronomy (CAASTRO)\\
$^5$Department of Physics, West Virginia University, 210E Hodges Hall, Morgantown WV 26506, USA\\
$^6$INAF-Osservatorio Astronomico di Cagliari, localit\'a Poggio dei Pini, strada 54, I-09012 Capoterra, Italy\\
$^7$Max Planck Institut f\"ur Radioastronomie, Auf dem H\"ugel 69, 53121 Bonn, Germany\\
$^8$Dipartimento di Fisica, Universit\'a degli Studi di Cagliari, Cittadella Universitaria, 09042 Monserrato (CA), Italy\\
$^9$University of Manchester, Jodrell Bank Centre for Astrophysics, Alan Turing Building, Manchester M13 9PL, UK
}

\date{}
\begin{document} 

\maketitle
\begin{abstract}
We report on the pulse-to-pulse energy distributions and phase-resolved modulation properties for catalogued pulsars in the southern High Time Resolution Universe intermediate-latitude survey. We selected the 315 pulsars detected in a single-pulse search of this survey, allowing a large sample unbiased regarding any rotational parameters of neutron stars.
%Approximately one quarter of these pulsars exhibited detectable nulling. 
We found that the energy distribution of many pulsars is well-described by a log-normal distribution, with few
%of those pulsars
deviating from a small range in log-normal scale and location parameters. Some pulsars exhibited multiple energy states corresponding to mode changes, and implying that some observed ``nulling'' may actually be a mode-change effect. PSR\,J1900--2600 was found to emit weakly in its previously-identified ``null'' state. We found evidence for another state-change effect in two pulsars, which show bimodality in their \emph{nulling time scales}; that is, they switch between a continuous-emission state and a single-pulse-emitting state.

Large modulation occurs in many pulsars across the full integrated profile, with increased sporadic bursts at leading and trailing sub-beam edges. Some of these high-energy outbursts may indicate the presence of ``giant pulse'' phenomena. We found no correlation with modulation and pulsar period, age, or other parameters.
%Some authors suggest pulsar single-pulse properties are linked to their emission class of conal or core emission, however our data lack the polarisation information needed to determine the geometry and hence emission class.
%However, lacking in information about core/conal profile type, we could not discern between several pulsar emission theories with predictions for such correlations. 
Finally, the deviation of integrated pulse energy from its average value was generally quite small, despite the significant phase-resolved modulation in some pulsars; we interpret this as tenuous evidence of energy regulation between distinct pulsar sub-beams.
\end{abstract}
\begin{keywords}
\end{keywords}

\section{Introduction}\label{sec:intro}
Radio pulsars have long been known to display a myriad of intrinsic amplitude modulation effects. Averaged over many rotations, most pulsars have a reproducible pulse shape, reflective of the long-term stability of their rotation and magnetism. In contrast, sequential rotations of a pulsar can differ considerably in pulse shape and intensity; ordered effects such as sub-pulse drift, mode changing, and nulling \citep[e.\,g.][]{subpulsedrift,originalnulling}, as well as stochastic pulse-to-pulse shape and intensity variations affect pulsars to varying degrees.
Other effects such as intense giant pulses \citep{crabpulsesdiscovery,crabpulsardiscovery} or ``giant micropulses'' (referencing their narrow structure, e.\,g.~\citealt{hiresvela}) occur in some pulsars at a limited phase range.

The energy distribution of radio pulses can provide a window into the state of pulsar plasma and the method of emission generation. There exist a great number of viable plasma-state models, a few of which predict pulse energy distributions; the most commonly-proposed predictions are of Gaussian, log-normal, and power-law distributions. \citet{SGT2} and \citet{SGT3}, and references therein, provide discussion on these models.
%(...insert words+citations here about linear/nonlinear plasma growth, and why we might expect gaussian, or log-normal, or powerlaw, or other distributions...).
Energy distributions have been examined in detail for only a few pulsars \citep{cognard,cjd01,cairnsotherpulsars}, resulting in the conclusion that those pulsars obey log-normal statistics. These analyses have substiantially contributed to the hypothesis that genuine ``giant pulses'' are generated separately from standard pulse generation; while ``giant pulse'' is sometimes used to refer to any single pulse of more than ten times the average intensity, studies have revealed giant pulses with power-law energy distributions, distinct from log-normal main pulse components
%, Crab pulsar, Vela, and PSR B1706--44
\citep{crabgiantpulsestats,kramerjohnstonstraten02,johnstonromani02}.
No survey targeting single-pulse energy distribution shapes or giant pulses in the general population has yet been performed.

Phase-resolved modulation analysis is likewise thought to be an indicator of radio emission's geometry and generation mechanism. \citet{weisberg86} first noted differences in modulation between core and conal-type pulse profiles, while \citet{jenetgil2003} derived theoretical predictions for anti-correlations between the modulation index (defined in \S\ref{sec:modparams}) and four ``complexity parameters,'' corresponding to four pulsar emission models. Their complexity parameters are: $a_1 = 5\dot P^{2/6}P^{-9/14}$, for the sparking gap model, $a_2 = (\dot P/P^3)^{0.5}$ for the continuous current outflow instabilities, $a_3 = (P\dot P)^{0.5}$ for surface magnetohydrodynamic wave instabilities, and $a_4 = (\dot P/P^5)^{0.5}$ for outer magnetospheric instabilities. The \citet{jenetgil2003} measurements of modulation index for a small sample of core-type profiles disfavoured the magnetohydrodynamic wave instability model.
The studies of \citet{patricketal06,patricketal07} surveyed ordered, longitude-resolved modulation in $\sim$190 pulsars at 21 and 92\,cm. Their large sample enabled them to test correlations with other neutron star properties. They determined that the modulation index is generally higher at lower frequencies, and noted a weak correlation between modulation index and age that is dampened at higher frequency. 

The study of single-pulse modulation in a large pulsar sample can also contribute to several practical questions, for instance: how common is giant-pulse emission, and are some ``giant pulses'' the manifestation of a broad log-normal energy distribution?,
Are the prospects of pulsar detection in other galaxies better for single-pulse or Fourier searches \citep[e.\,g.][]{johnstonromani03,mclaughlincordes03}?
% we can speculate on this with R, mod distribution, and maybe <E> or SP detection reporting
%How latitude-dependent is modulation, and what does this imply for emission geometries?
Quantification of pulsars' modulation will also aid in understanding the physical makeup of ``rotating radio transients'' (RRATs; \citealt{rrats}). Energy distributions in bright, individual RRATs show that some appear to be pulsars with extremely high ($\gg$95\%) nulling fractions \citep[e.\,g.][]{sbsmb,htruSP,millerdists}. However, an unknown fraction of RRATs may be distant pulsars with extremely broad energy distributions, such that only their brightest, infrequent pulses are detectable \citep{modpulsar}. The distinction between these two cases will be critical in quantifying RRATs' potentially overwhelming contribution to galactic pulsar populations \citep{evanbirthrates}, however the general pulsar population's intrinsic energy distributions have not yet been extensively studied.

The High Time Resolution Universe survey recently completed its southern intermediate latitude survey (``HTRU med-lat'') of galactic latitudes $|b| < 15^\circ$ and longitudes $-120^\circ<l< 30^\circ$ for pulsars \citep{htru1} and single pulses \citep{htruSP}. Single pulses from known pulsars were detected at rates that vastly improve on previous surveys of the same region, testifying to the increased sensitivity of the high dynamic range, frequency resolution and time resolution of a new digital search backend on Parkes Radiotelescope \citep[described in][]{htru1}.
%This represents a significant fraction of pulsars detectable in single pulses, and the success rate is likely due to a combination of the high dynamic range of the 2-bit system, and the improved sensitivity to narrow and highly-dispersed pulses afforded by our high frequency and time resolution.

In this paper, we study the modulation properties of all med-lat pulsars with detectable single pulses, using the relatively unbiased, single-pulse flux-limited sample provided by the HTRU med-lat survey. We focus here on studies that can be performed within the survey's 9-minute observations, pursuing pulse intensity distribution statistics and the measurement of basic pulse-to-pulse modulation properties. In \S\ref{sec:sample} we describe our sample selection, and \S\ref{sec:methods} describes our analysis methods. In \S\ref{sec:edists} and \S\ref{sec:modstats}, we describe the results of our energy distribution analysis and modulation analysis, respectively, and provide discussion of the results. \S\ref{sec:remaining} reviews other science aspects addressed by our analysis. \S\ref{sec:conclusions} summarises our findings.

%homogeneously-observed, high-quality data for a large fraction of pulsars makes the HTRU survey uniquely
%why it is good for this analysis and what it adds to previous studies (e.\,g.~a minimally-biased sample, larger number of studied objects, and over a somewhat wider range in physical pulsar parameters). HTRU did better than other surveys at finding PSRs with single pulses, because of backend I suppose (better dynamic range and narrow-pulse sensitivity---cite htruSP).
%Generic and brief introduction about large surveys of single-pulse properties of pulsars, nulling, modulation, and related effects. Brief mention on why HTRU is good for this analysis and what it adds to previous studies (e.\,g.~a minimally-biased sample, larger-number statistics, maybe a wider range in physical pulsar parameters).
%
%In this paper we perform statistical tests on all PSRs with detectable single pulses in the HTRU survey.

\section{Data and Pulsar Sample}\label{sec:sample}
Our data is made up of HTRU med-lat survey observations. This survey had 64\,$\mu$s sampling, and a bandwidth of 340\,MHz is divided into 870 frequency channels, centred on 1352\,MHz. Two polarisation channels are summed prior to data recording, and data is digitised using 2\,bits. The system temperature was $23$\,K.
%The med-lat survey covered galactic latitudes $|b| < 15^\circ$ and longitudes $-120^\circ<l< 30^\circ$.

\subsection{Determination of pulsar sample}\label{sec:psrsample}
The initial pulsar set included all pulsars in the med-lat survey region as queried through the online ATNF Pulsar Database ({\sc psrcat}).\footnote{Originally published by \citet{psrcat}, available at http://www.atnf.csiro.au/research/pulsar/psrcat/} We selected the observation of smallest angular distance within 0.25 degrees to each pulsar, yielding 1159 observations near 1113 pulsars (some had multiple observations at roughly equal distance to the pulsar).
We scrutinised the HTRU Fourier and single-pulse search results for each observation (as described in \citealt{htru1} and \citealt{htruSP}, respectively) to determine the pulsar's detectability. Single pulses were ``detected'' if a pulse peaking near the pulsar's dispersion measure (DM) exceeded a significance of 6, and was confirmed by inspection of the data. 411 pulsars were not detected by single-pulse or Fourier analysis,\footnote{These non-detections were investigated, and typically found to be due to strong interference, scintillation, or insufficient integration time (i.\,e. the faint objects discovered by the 35-\,minute Parkes Multibeam Survey pointings, \citealt{pksmb})} and 16 were detected only through single-pulse analysis.
%, likely due to scintillation, the presence of strong interference, or insufficient sensitivity.
Of the Fourier-detected pulsars, 45\%
%315/(1113-411)
had at least one detected single pulse. It is the 315 pulsars with detected single pulses that we analyse in this study.
%This represents a significant fraction of pulsars detectable in single pulses, and the success rate is likely due to a combination of the high dynamic range of the 2-bit system, and the improved sensitivity to narrow and highly-dispersed pulses afforded by our high frequency and time resolution.

%Figure \ref{fig:ppdot} illustrates the make-up of our sample.
Our sample is not isolated in period-period derivative phase space, consistent with previous studies \citep[e.\,g.][]{sbsmb}. We explore the full range of pulsars' magnetic field strength ($B$), energy derivative ($\dot E$), period ($P$), period derivative ($\dot P$), and characteristic age ($\tau_{\rm c}$), giving us acute sensitivity to any dependence of modulation effects on these parameters. Our sample includes two millisecond pulsars (PSRs J1439--5501 and J1744--1134) and one radio magnetar (PSR\,J1622--4950; \citealt{htrumagnetar}).

%\begin{figure}
%\centering
%\includegraphics[angle=270,width=0.48\textwidth,trim=0mm 0mm 0mm 0mm, clip]{figs/knownSPppdot.png}
%\caption{MAKE THIS A ZOOM-IN ON MAIN PSR ISLAND? A period-period derivative diagram indicating (.....) %Detected by SPs but not by Fourier search/folding---these indicative of the HTRU ``RRAT sample'' I suppose. (BUT NOTE IN THE FIGURE) that the HTRU ``RRAT sample'' also consists of normal pulsars and drive home the fact that RRAT is a detection label!
%Show all the PSRs we saw, and those with $\se$ above whatever value we choose and reference Sections \ref{sec:psrsample} and \ref{sec:fluxenergy}. Also specially mark the things only detected by SPs and not by the Fourier pipeline and make strong derisive statements about RRATs.
%}\label{fig:ppdot}
%\end{figure}

\subsection{Pulse Stacks and Data Configuration}
We dedispersed each observation and resampled the resulting time series to break it into integrations of duration equal to the pulsar's rotational period. We used DMs and periods as predicted by {\sc psrcat} ephemerides. In some cases, the observed period did not match the ephemeris prediction. For these we used the rotational period measured in our observation. 
Each integration consisted of $1024$ phase bins, or in some cases integer divisors of two where needed
%(for pulsars with $P\lesssim 65.5$\,ms)
to ensure the size of one bin equaled or exceeded the original data's sampling time. This caused slightly degraded longitude resolution for short-period pulsars.

\begin{figure*}
\centering
\begin{tabular}{cc}
%\subfigure[PSR\,J1808--2057 energy distribution]{
\includegraphics[trim=9mm 0mm 0mm 0mm, clip,angle=270,width=0.47\textwidth]{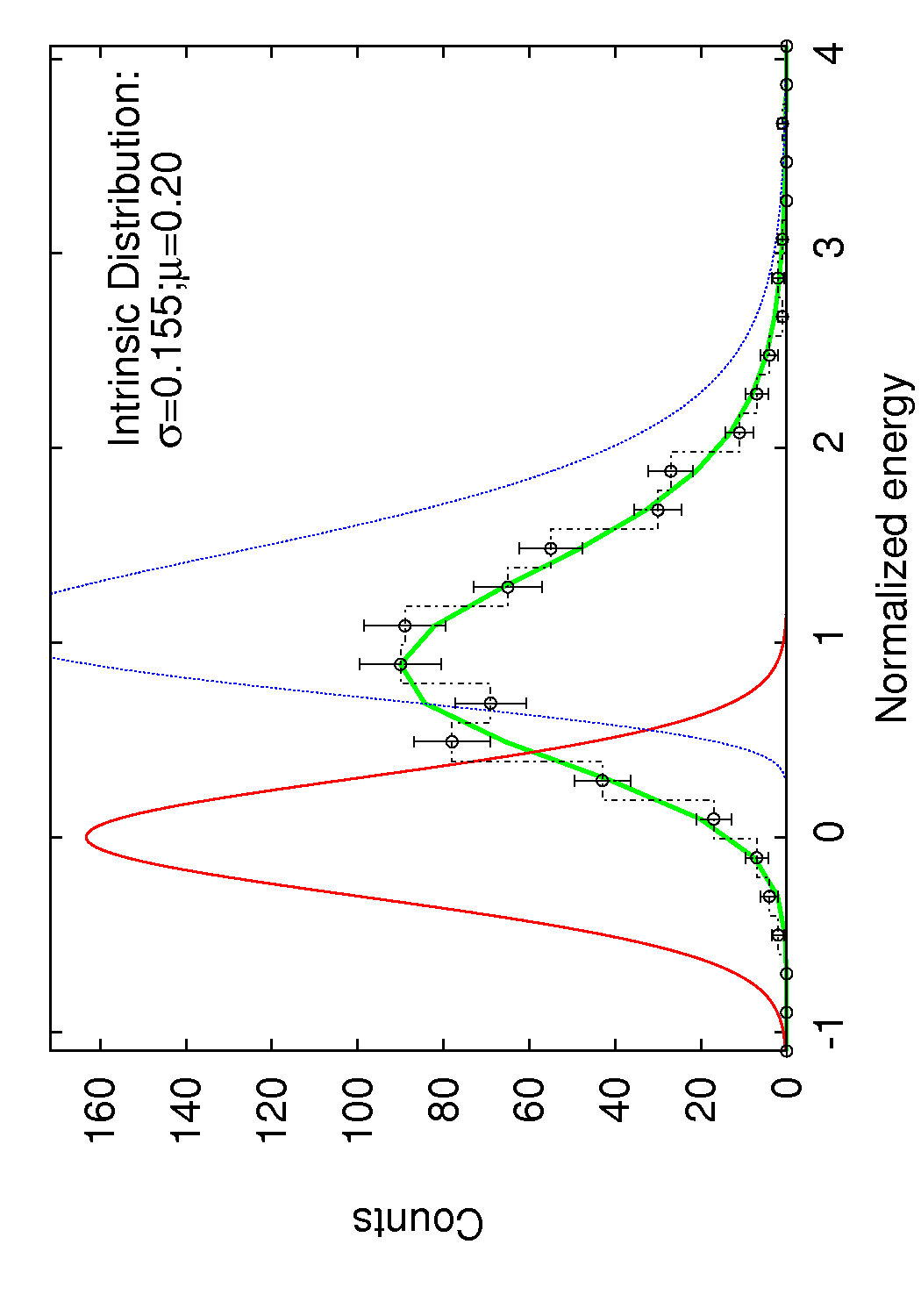}&%\vspace{1mm}&
%}
%\subfigure[PSR\,J1428--5530 energy distribution]{
\includegraphics[trim=9mm 0mm 0mm 0mm, clip,angle=270,width=0.47\textwidth]{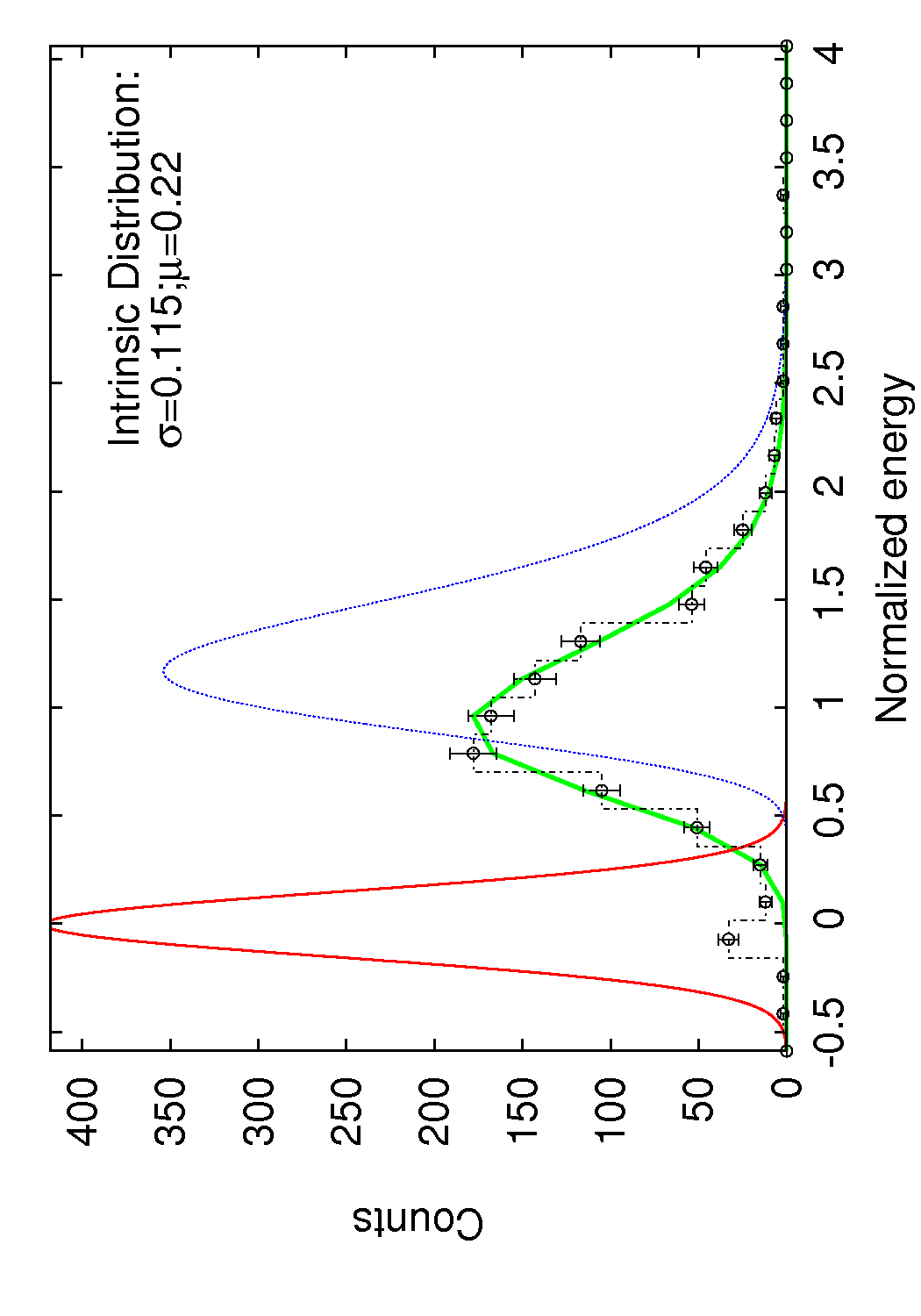} \vspace{-3mm}\\
%}
\subfigure[PSR\,J1808--2057]{
\includegraphics[angle=180,width=0.47\textwidth,trim=3.5mm 20mm 0mm 14mm, clip]{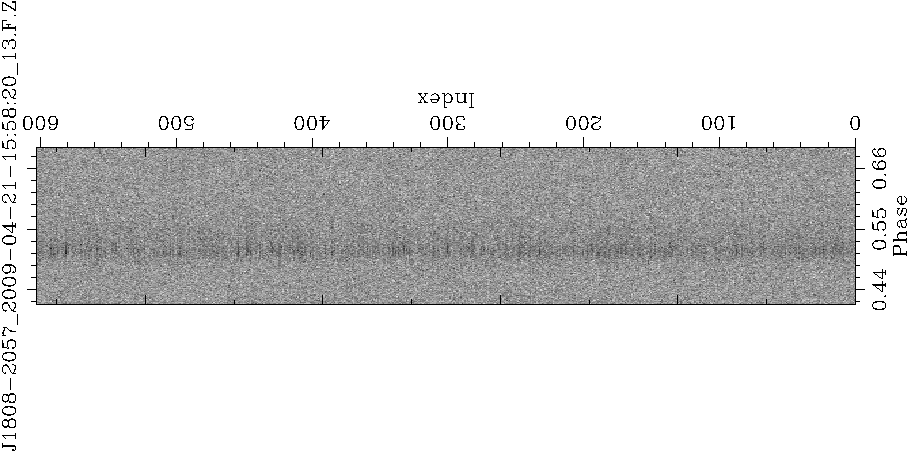}
} &
\subfigure[PSR\,J1428--5530]{
\includegraphics[angle=180,width=0.47\textwidth,trim=3.5mm 20mm 0mm 14mm, clip]{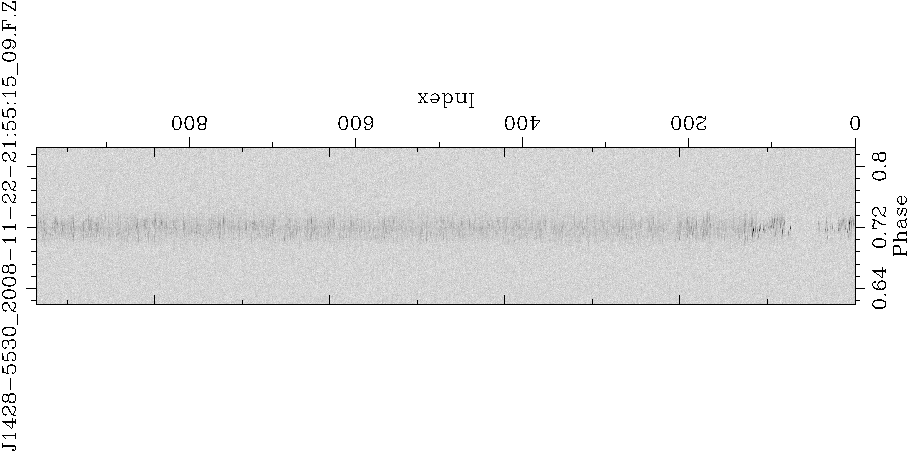}\label{fig:nulldist}
}
%\subfigure[]{
%\includegraphics[width=0.43\textwidth,trim=0mm 0mm 0mm 0mm, clip]{figs/}
%}
\end{tabular}
\vspace{-5mm}
\caption{Two examples of pulsars with log-normal energy distributions. The upper plots show the energy distribution of on-pulse data (black dash-dot histogram with points and error bars), the log-normal energy model fits (thick green line), the off-pulse noise model (thin red line), and the intrinsic energy distribution (blue dotted line). Errors shown are the square root of the number in each bin. The lower plots show the corresponding pulse stacks for each pulsar. PSR\,J1428--5530, in $(b)$, has a brief nulling episode from pulse indexes $\sim$50-80 that is visible, distinct from the log-normal distribution for non-null pulses. A Figure showing all pulse stacks, energy distributions, and phase-resolved modulation is available online (see Appendix \ref{sec:appendix}).}\label{fig:distplots}
\end{figure*}

Throughout this analysis we refer to ``pulse stacks,'' which are the observed power represented as a function of pulse phase and number (indexed from the observation's start), as shown in the lower panels of Figure \ref{fig:distplots}.

%\begin{figure*}
%\centering
%\includegraphics[trim=3.5mm 0mm 0mm 0mm, clip,width=0.65\textwidth,angle=180]{figs/teststack.png}
%\end{figure*}

\section{Methods and Analysis}\label{sec:methods}
\subsection{Flux and energy measurements}\label{sec:fluxenergy}

Normalised energy, $\enorm$, was calculated using the standard method for determining pulse energy distributions (e.\,g.~\citealt{ritchings76}, \citealt{biggs92}). In each observation, we defined on-pulse windows of size $N_{\rm on}$ bins, the position and width of which were determined by inspection of the integrated pulsar profile. Where the pulsar duty cycle was less than 0.5, we chose an off-window also of size $N_{\rm on}$ to determine the off-pulse energy. All bins not part of the on-window were used to estimate the integration's per-bin standard deviation, and to remove a baseline from all bins.
We did not divide integrations into shorter analysis blocks (as in \citealt{ritchings76}), as interstellar scintillation at our observing frequency for each pulsar was expected, and observed, to be minimal based on the NE2001 galactic electron density and scintillation model \citep{ne2001}. Furthermore, it was realised that block analysis mutes the modulation induced by intermediate-timescale nulling and mode-changing in some pulsars. The normalised on-pulse and off-pulse energies ($\eon$ and $\eoff$, respectively) were calculated for each stellar rotation by integrating the energy in the on- and off-window bins, respectively, then dividing by the mean on-pulse energy of all integrations.
The single-pulse energy significance, $\se$, is given by $\se=\eon/(\sigma\sqrt{N_{\rm on}})$, where $\sigma$ is the standard deviation of the off-pulse energy. 
%We used the average of this value over all sub-integrations in each observation to define the ``bright source sample'' noted in Sec.\,\ref{sec:psrsample}. 
% consisting of pulsars with \mbox{$\langle m_E\rangle>(5-or-6)$}. In other words, pulsars for which on average each single-pulse energy was measured to a significance of more than ($5-or-6$). This sample is used for e.\,g. modulation statistics whose results are sensitive to the presence of noise.

\subsection{Energy distribution tests}\label{sec:etests}
We performed analysis of pulse energy distributions to assess whether the probability distribution function of pulse energy is well-fit to either a log-normal or Gaussian distribution, and if so, whether pulsars share typical distribution parameters. We analysed the $\eon$ distributions by constructing histograms of $\eon$ and $\eoff$ in 25 fixed-size bins over the full range of detected normalised energy for each available observation. We modelled each observation's noise with a Gaussian of the same mean and standard deviation as the off-pulse distribution.

We then performed a least-squares minimisation of the data fit by a model for the intrinsic on-pulse energy distribution convolved with noise. The test distributions were defined by Gaussian and log-normal probability density functions. For the Gaussian case, we tested a grid of probable values in the range $0.02<\sg<1.10, 0.2<\mg<4.0$, where $\sg$ and $\mg$ are the standard deviation and mean of the distribution, respectively. In the log-normal case, we tested over scale and location parameters in the range $0.02<\sln<1.10, -2.0<\mln<2.0$, where these parameters are defined in the probability distribution as:
\begin{equation}
	P(E) = \frac{1}{E\sln\sqrt{2\pi}}~{\rm exp}\bigg[\frac{-({\rm log}_{10}(E) - \mln)}{2\sln^2}\bigg]
\end{equation}
We sampled each test distribution at equal bin size and range as the data, then convolved it with the Gaussian noise model. A least-squares fit was then computed between the noise-convolved model and the data.

The goodness of fit of the best-fit distribution was quantified by a $\chi^2$ analysis, using only bins where the value of the convolved model in the bin was greater than five.\footnote{Note that the $\chi^2$ value we use is defined using \mbox{$\chi^2=\sum[({\rm data\,value} - {\rm model\,value})^2/({\rm model\,value})] $}, which avoids the use of ill-defined errors on our distributions. This is not expected to introduce a bias in the measured goodness-of-fit probability or distribution parameters, because the initial fit was performed using a least-squares minimisation that took the full distribution into account.} We took the degrees of freedom to be the number of valid bins minus 3, and a goodness-of-fit probability was calculated from the $\chi^2$ cumulative distribution function and the fit's $\chi^2$ value. Probabilities were calculated for both the logarithmic and Gaussian cases (giving $\pln$ and $\pg$, respectively). Finally, the best-fit convolved Gaussian and log-normal models were overlaid on the data (e.\,g.~Fig. \ref{fig:distplots}) and inspected by eye to aid in classification of the energy distributions. The results of this analysis are described and discussed in Sec.\,\ref{sec:edists}.

\subsection{Recognition of pulse nulling}
We performed an inspection of both pulse stacks and pulse energy distributions to determine whether nulls were either not present, or clearly present, in each observation. In Table \ref{table:stats}, we indicate for each pulsar whether no null pulses were observed (marked by {\bf N}, indicating no pulses occurred at a zero energy state), or whether a peak at zero energy was discernible from a distinct on-pulse distribution (marked by {\bf Y}). For the remaining pulsars, we could not distinguish the presence or non-presence of nulls without further analysis, which will be performed in the future for all pulsars observed in the HTRU med-lat survey. We could distinguish 31 pulsars ($\sim$10\% of the full sample) with no observed nulls, and 69 pulsars ($\sim$22\% of the sample) with a null state. The remaining pulsars in the sample were not sufficiently bright to distinguish whether they were nulling or not.
 %NULLS: give more detail about nulling recognition? visual nulls vs. actual nulls (as determined by pulse energy dist and subint nulls and literature confirmation of nulling
%As our energy distribution tests inspect only unimodal distributions, and energy distribution of nulling pulsars is not unimodal, we cannot report reliable fits as described above for those pulsars. 

\subsection{Parameterisation of modulation}\label{sec:modparams}
We quantified the longitude-resolved modulation in each pulsar by computing two values for each bin of a pulse stack. The observed ``modulation index'' is defined as $m_{\rm obs,j} = \sigma_j/\mu_j$, where $\sigma_j$ and $\mu_j$ are the standard deviation and the mean across the whole observation, respectively, of the $j$th bin. Interstellar scintillation can induce signal in $m_{\rm obs,j}$ of $m_{\rm ISM} = (1+\eta B/\delta b)^{-1/2}$, where $B=340$\,MHz is the receiver bandwidth, $\eta$ is a filling factor, and $\delta b$ is the scintillation bandwidth of the pulsar. We determined $\delta b$ from the \citealt{ne2001} galactic electron density model and scaled the value to the centre HTRU survey observing frequency of 1.352\,GHz assuming $\delta b\propto f^4$. The induced $m_{\rm ISM}$ value was inferred using the prescription of \citet{jenetgil2003}, where setting $\eta=0.18$, the intrinsic modulation index is:
\begin{equation}
	m = \bigg(\frac{m_{\rm obs}^2-m_{\rm ISM}^2}{m_{\rm ISM}^2+1}\bigg)^{1/2}~.
\end{equation}
The modulation index is most sensitive to persistent oscillations in a pulsar's signal, e.\,g. as would be caused by sub-pulse drift or mode-changing on timescales much less than the observation time. This parameter has poor accuracy for observations of low integrated signal to noise, for instance it is undefined off-pulse, and is insensitive to non-persistent modulation like sporadic or infrequent outbursts.

%\footnotetext[4]{Describe how to access online Figure in this footnote after discussion with editor. Presently it is available at http://dl.dropbox.com/u/22076931/supplementary\_material.pdf}

To identify sparse modulated emission (on- and off-pulse), we use the $R$ modulation statistic introduced by \citet{hiresvela}. They define the $R$-parameter as $\rj = (M_{\rm j} - \mu_{\rm j})/\sigma_{\rm j}$, again computed in the $j$th bin of each observation, where $M_{\rm j}$ is the maximum value observed in that bin. Given that the per-bin statistics are (in the absence of pulsar signal or interference) Gaussian-distributed, even off-pulse regions are expected to exhibit $\rj$ values consistent with a Gaussian distribution. This and its dependence on the significance of mean single pulse brightness render it difficult to use as an absolute comparative modulation statistic between pulsars, however it is ideal for identifying the presence of giant pulses, and other extreme phase-dependent, sparse modulation or significantly non-Gaussian behaviour.
%To correct for this tie between $\rj$ and number of observed rotations, we use a modified $r$ statistic, where $\rj = \rj-\langle \rj\rangle_{\rm off}$, where $\langle \rj\rangle_{\rm off}$ is the off-pulse average $\rj$ value.
We consider a measurement of $\rj$ ``significant'' if the bin's value minus the off-pulse mean is more than four times the standard deviation of the $\rj$ values in the off-pulse window.
%While sensitive to
%The $R$ parameter is sensitive to extreme outliers in the energy distribution at a given pulse phase, making it ideal for accenting e.\,g. infrequent giant pulses or latitudes with extended energy tails.
%By the way this parameter is defined, the ``average'' off-pulse value of $\rj$ will depend on the number of neutron star rotations in an observation (the off-pulse $\langle R\rangle$ will increase with the number of rotations due solely to Gaussian noise statistics).

%These two modulation parameters are in numerous ways complimentary to one another. 
%---$>$ {\sc Note then how $\rj$ is dependent on S/N (becomes less at higher S/N) . Go briefly into the difference between these two modulation parameters---explicitly note the different modulation effects they're sensitive to}.

\begin{table*}
{\scriptsize
\begin{centering}
  \caption{Here we report the numerical results of the energy distribution and modulation analysis. Columns are: 1) PSR name (J2000); %!!! Marked with * only detected in SP and not FFT);
2) Number of pulses detected in blind single pulse search, and total number of rotations in the observation; 3) Integrated (Fourier) signal-to-noise ratio (S/N);
4) Signal-to-noise ratio of the brightest single pulse detected in the blind single pulse search; 
5) Average single pulse energy significance; 
6) Maximum $\rj$ value, where significant; 
7) Minimum on-pulse, phase-resolved modulation index, where significant; 
8) Indication of whether pulse appears to be nulling (Y) or had no zero-energy pulses (N); 
9) Energy distribution classification; 
10--12) The probability and fit values associated with the best-fit log-normal energy distribution;
13--15) The probability and fit values associated with the best-fit Gaussian energy distribution.\vspace{1mm} ~~~~~~~~~~~~~~~~~~~~~~~~~~~~~~~~~~~~~~~~~~~~~~~~~~~~~~~~~~~~~~~~~~~~~~~~~~~~~~~~~~~~~~~~~~~~~~~~~~~~~~~~~~~~~~~~~~~~~~~~~~~~~~~~~~~~~~~~~~~~~~~~~~~~~~~~~~~~~~~~~  *~Many single pulses from the Vela pulsar (PSR\,J0835--4510) saturated the observing instrumentation. This disrupted the observed integrated and single-pulse S/N, and the modulation parameters at phases near the pulse peak. The modulation parameters are not reported here but Vela's modulation profile can be viewed in the online figure (see Appendix \ref{sec:appendix}).}
\label{table:stats}
%{\bf PUT $\mln, \mu_{G}$ HERE OR NO? NOT GENERALLY USEFUL BUT MAYBE...}
\begin{tabular}{ccccccccccccccc}
\hline
(1)               & (2)        & (3)          & (4)       & (5)                   & (6)              & (7)            & (8)              & (9)             & (10) & (11) & (12) & (13) & (14) & (15) \\
{\bf PSR}& & {\bf S/N}&{\bf S/N} &{\bf }     &{\bf Max.}  &{\bf Min.}    & {\bf }          &{\bf Dist.}   &\\
%\multicolumn{2}{c}{\bf best-fit} &{\bf etc.} &\\
{\bf Jname} & Npulses & {\bf (int)}&{\bf (SP)}&{\bf $\se$}&{\bf $R_{\rm j}$}&{\bf $m_{\rm j}$}& {\bf Null?}& {\bf class} & {\bf $\pln$}& {\bf $\sln$}& {\bf $\mln$}& {\bf $\pg$}&{\bf $\sg$} &{\bf $\mg$}\\ 
\hline

 J0726--2612 &       17/163 &   18.4 &  35.4 &   1.2 &   9.3 &    -- &   Y &  -- & 0.2363 & 0.05 & 1.02 & 0.0023 & 0.30 & 2.86 \\
 J0738--4042 &    1469/1482 & 2645.5 &  47.1 &  53.3 &  11.1 &    0.3523 &   N &   G & 0.1387 & 0.07 & 0.12 & 0.7804 & 0.21 & 1.15 \\
 J0742--2822 &    3306/3341 & 1303.5 &  35.4 &  19.9 &   6.3 &    0.3353 &   N &   O & 0.0119 & 0.07 & 0.10 & 0.0000 & 0.18 & 1.10 \\
 J0745--5353 &     126/2625 &  163.6 &  15.6 &   2.7 &   5.5 &    1.8748 &  -- &   O & 0.0017 & 0.10 & 0.27 & 0.0000 & 0.30 & 1.34 \\
 J0809--4753 &      12/1024 &  101.8 &   8.0 &   2.7 &    -- &    1.0487 &  -- &   L & 0.9372 & 0.09 & 0.15 & 0.2798 & 0.24 & 1.20 \\
 J0818--3232 &       49/251 &   64.1 &  13.7 &   2.9 &   5.6 &    -- &  -- &   L & 0.9289 & 0.17 & 0.17 & 0.1499 & 0.44 & 1.20 \\
 J0820--4114 &       1/1030 &  103.0 &   6.5 &   3.4 &    -- &    2.7182 &  -- &   O & 0.0000 & 0.17 & 0.32 & 0.3852 & 0.53 & 1.39 \\
 J0828--3417 &       10/296 &   14.5 &  41.4 &   1.0 &  11.5 &    -- &   Y &  -- & 0.0021 & 0.35 & 1.00 & 0.0000 & 1.09 & 2.48 \\
 J0831--4406 &       1/1810 &   32.1 &   6.4 &   0.6 &    -- &    -- &  -- &   L & 0.9425 & 0.18 & 0.62 & 0.5718 & 0.87 & 1.96 \\
 J0835--3707 &       6/1023 &   31.6 &  13.2 &   0.7 &   9.7 &    -- &  -- &   U & 0.9934 & 0.17 & 0.67 & 0.9497 & 0.87 & 2.05 \\
 J0835--4510 &    6206/6259 & 18957.3 &  22.1 & 146.1 &  17.2 &    0.0649 &   N &   O & 0.0000 & 0.05 & 0.07 & 0.0000 & 0.11 & 1.05 \\
 J0837--4135 &      726/737 & 2139.3 & 149.0 &  45.1 &  12.6 &    0.5839 &  -- &   M & 0.0063 & 0.14 & 0.20 & 0.0000 & 0.36 & 1.25 \\
 J0840--5332 &        2/779 &   63.1 &   6.7 &   1.5 &    -- &    -- &  -- &   L & 0.8239 & 0.14 & 0.40 & 0.1862 & 0.51 & 1.53 \\
 J0842--4851 &        2/874 &   44.1 &   8.5 &   1.0 &   5.6 &    -- &  -- &   L & 0.8862 & 0.15 & 0.65 & 0.1996 & 0.72 & 1.96 \\
 J0846--3533 &      219/490 &  138.7 &  18.9 &   5.3 &   5.7 &    0.6191 &  -- &   L & 0.9933 & 0.10 & 0.20 & 0.7240 & 0.29 & 1.25 \\
 J0855--3331 &       57/443 &   65.4 &  16.4 &   2.2 &   6.0 &    -- &   Y &  -- & 0.0017 & 0.26 & 0.42 & 0.4117 & 0.94 & 1.48 \\
 J0902--6325 &        1/842 &   36.7 &   6.2 &   1.1 &    -- &    -- &  -- &   O & 0.0548 & 0.06 & 0.55 & 0.0153 & 0.28 & 1.72 \\
 J0904--4246 &        5/564 &   29.7 &   9.7 &   1.0 &   5.3 &    -- &  -- &   L & 0.9844 & 0.19 & 0.45 & 0.6663 & 0.72 & 1.63 \\
 J0907--5157 &    1170/2208 &  456.0 &  29.0 &   7.2 &  10.6 &    0.8975 &  -- &   O & 0.0048 & 0.15 & 0.12 & 0.0000 & 0.38 & 1.10 \\
 J0908--4913 &    4619/5224 &  102.8 &  40.3 &   7.5 &   7.2 &    0.4716 &  -- &   O & 0.0000 & 0.09 & 0.15 & 0.0008 & 0.22 & 1.15 \\
         (i) &           -- &     -- &    -- &   0.6 &   7.2 &    0.4807 &   N &   O & 0.0002 & 0.14 & 0.87 & 0.3194 & 0.80 & 2.43 \\
         (m) &           -- &     -- &    -- &   8.7 &    -- &    0.4716 &   N &   O & 0.0000 & 0.09 & 0.17 & 0.0000 & 0.25 & 1.20 \\
 J0922--4949 &        2/591 &   24.6 &   7.9 &   0.8 &   5.5 &    -- &  -- &   O & 0.0251 & 0.04 & 0.65 & 0.0032 & 0.13 & 1.91 \\
 J0924--5302 &        1/614 &   48.0 &   6.7 &   1.5 &   4.4 &    -- &  -- &   L & 0.9909 & 0.09 & 0.35 & 0.6844 & 0.26 & 1.44 \\
 J0924--5814 &       36/724 &   69.5 &   9.2 &   2.6 &   5.0 &    -- &  -- &   O & 0.6873 & 0.14 & 0.35 & 0.0000 & 0.41 & 1.44 \\
 J0934--5249 &      349/385 &  221.7 &  51.9 &   8.5 &  11.4 &    0.8599 &   Y & (U) & 1.0000 & 0.10 & 0.20 & 0.9613 & 0.28 & 1.25 \\
 J0942--5552 &      445/842 &  327.2 &  41.9 &   9.0 &   4.8 &    0.8019 &  -- &   M & 0.0087 & 0.22 & 0.15 & 0.0000 & 0.53 & 1.10 \\
 J0942--5657 &      198/682 &  121.5 &  12.7 &   2.9 &   4.7 &    0.6559 &  -- &   L & 0.9973 & 0.10 & 0.27 & 0.6542 & 0.33 & 1.29 \\
 J0945--4833 &       1/1687 &   32.0 &   6.6 &   0.6 &    -- &    -- &  -- &   U & 0.9988 & 0.17 & 0.65 & 0.9885 & 0.83 & 2.01 \\
 J0955--5304 &       12/652 &   40.4 &   9.0 &   1.3 &   5.2 &    -- &  -- &   U & 0.9993 & 0.21 & 0.37 & 0.9230 & 0.79 & 1.44 \\
 J1001--5507 &      376/386 &  410.8 &  45.4 &  14.5 &   5.2 &    0.4926 &   N &   U & 0.8644 & 0.13 & 0.15 & 0.9681 & 0.34 & 1.15 \\
 J1001--5559 &        1/334 &   31.2 &   6.2 &   1.4 &    -- &    -- &  -- &   O & 0.0523 & 0.02 & 0.45 & 0.0036 & 0.09 & 1.58 \\
 J1001--5939 &        18/70 &   21.9 &  15.8 &   1.0 &   4.1 &    -- &  -- &   O & 0.3614 & 1.07 & -0.85 & 0.2075 & 0.17 & 1.05 \\
 J1003--4747 &       1/1823 &   47.6 &   6.5 &   0.9 &    -- &    -- &  -- &   O & 0.1650 & 0.22 & 0.35 & 0.1353 & 0.86 & 1.44 \\
 J1012--5830 &        1/262 &    6.0 &   6.2 &   0.3 &    -- &    -- &   Y &  -- & 0.0365 & 0.04 & 1.07 & 0.0010 & 0.49 & 2.95 \\
 J1012--5857 &      150/683 &  108.0 &  26.1 &   3.0 &   6.4 &    1.0507 &  -- &   O & 0.6336 & 0.14 & 0.30 & 0.0001 & 0.40 & 1.34 \\
 J1013--5934 &      51/1268 &   78.3 &  11.9 &   2.1 &   5.0 &    -- &  -- &   L & 0.9548 & 0.14 & 0.30 & 0.4019 & 0.43 & 1.39 \\
 J1016--5345 &       61/729 &   76.5 &  13.8 &   2.0 &   5.7 &    -- &  -- &   L & 0.9809 & 0.21 & 0.17 & 0.6692 & 0.59 & 1.20 \\
 J1017--5621 &     157/1105 &   84.0 &  17.4 &   1.9 &   7.3 &    -- &  -- &   L & 0.8495 & 0.15 & 0.32 & 0.0009 & 0.45 & 1.39 \\
 J1020--5921 &        4/448 &   22.6 &   9.5 &   0.8 &   6.8 &    -- &  -- &   L & 0.9745 & 0.25 & 0.37 & 0.7425 & 1.03 & 1.44 \\
 J1032--5206 &        2/231 &   24.4 &   7.3 &   1.3 &   5.1 &    -- &  -- &   L & 0.9552 & 0.09 & 0.45 & 0.4127 & 0.25 & 1.58 \\
 J1032--5911 &       4/1214 &   50.6 &   7.5 &   1.2 &    -- &    -- &  -- &   O & 0.0694 & 0.23 & 0.45 & 0.6103 & 0.98 & 1.53 \\
 J1036--4926 &      65/1096 &   78.5 &  12.5 &   1.8 &   6.3 &    -- &  -- &   L & 0.7617 & 0.09 & 0.35 & 0.0531 & 0.28 & 1.44 \\
 J1038--5831 &        2/846 &   30.7 &   7.2 &   0.8 &   4.7 &    -- &  -- &   M & 0.2974 & 0.11 & 0.45 & 0.0517 & 0.41 & 1.63 \\
 J1042--5521 &       31/479 &   76.4 &  10.3 &   2.7 &   4.5 &    -- &  -- &   L & 0.9559 & 0.14 & 0.17 & 0.3674 & 0.40 & 1.20 \\
 J1043--6116 &      68/1930 &   44.8 &  13.6 &   0.7 &   4.8 &    -- &  -- &   O & 0.0076 & 0.18 & 0.70 & 0.0009 & 0.92 & 2.05 \\
 J1046--5813 &       3/1525 &   67.4 &   7.2 &   1.4 &    -- &    -- &  -- &   U & 0.9953 & 0.11 & 0.42 & 0.8222 & 0.40 & 1.58 \\
 J1047--6709 &     173/2836 &   76.8 & 139.7 &   1.2 &  19.5 &    -- &  -- &   O & 0.0000 & 0.11 & 1.60 & 0.0000 & 1.09 & 3.95 \\
 J1048--5832 &    2235/4561 &  408.5 &  41.0 &   4.8 &  13.1 &    1.1688 &  -- &   M & 0.0000 & 0.33 & 0.07 & 0.0000 & 0.64 & 0.96 \\
 J1049--5833 &       54/239 &   40.5 &  11.0 &   2.0 &   5.8 &    -- &   Y &  -- & 0.0004 & 0.35 & 0.35 & 0.0000 & 1.09 & 1.29 \\
 J1055--6905 &       22/188 &   33.2 &  10.2 &   1.8 &   4.6 &    -- &   Y &  -- & 0.2285 & 0.31 & 0.40 & 0.0061 & 1.09 & 1.34 \\
 J1056--6258 &    1225/1326 &  974.0 &  39.9 &  22.3 &   9.5 &    0.4494 &   N &   O & 0.0000 & 0.06 & 0.12 & 0.0692 & 0.17 & 1.15 \\
 J1057--5226 &     237/2847 &   57.9 &  34.2 &   1.7 &  14.8 &    -- &  -- &   O & 0.0217 & 0.15 & 0.60 & 0.3086 & 0.63 & 1.86 \\
         (i) &           -- &     -- &    -- &   1.0 &    -- &    -- &   N &   O & 0.0284 & 0.13 & 0.87 & 0.0000 & 0.72 & 2.48 \\
         (m) &           -- &     -- &    -- &   1.1 &  14.8 &    -- &   N &   O & 0.0058 & 0.30 & 0.10 & 0.0001 & 0.94 & 1.10 \\
 J1059--5742 &       91/454 &   75.1 &  17.1 &   2.6 &   8.8 &    -- &   Y & (M) & 0.8808 & 0.14 & 0.22 & 0.0728 & 0.38 & 1.25 \\
 J1104--6103 &       5/2008 &   18.1 &  12.6 &   0.3 &   7.1 &    -- &  -- &   O & 0.1278 & 0.26 & 1.02 & 0.0000 & 1.09 & 2.95 \\
 J1106--6438 &        2/200 &   18.8 &   6.7 &   0.9 &    -- &    -- &  -- &   U & 0.9981 & 0.10 & 0.25 & 0.8714 & 0.38 & 1.29 \\
 J1107--5907 &       1/2220 &    4.5 &   6.2 &   0.0 &    -- &    -- &   Y &  -- & 0.0000 & 1.07 & 1.97 & 0.0000 & 0.24 & 3.76 \\
 J1110--5637 &     175/1006 &  153.6 &  18.4 &   4.3 &   6.6 &    0.9004 &  -- &   O & 0.7491 & 0.10 & 0.17 & 0.0950 & 0.28 & 1.20 \\
 J1112--6926 &        9/679 &   68.9 &   9.7 &   2.2 &   5.2 &    -- &  -- &   L & 0.8444 & 0.09 & 0.27 & 0.3124 & 0.26 & 1.34 \\
 J1114--6100 &       94/639 &  109.2 &  14.6 &   3.8 &   4.8 &    1.1906 &  -- &   L & 0.9568 & 0.22 & 0.12 & 0.0096 & 0.53 & 1.10 \\
 J1117--6154 &       7/1099 &   42.1 &   9.4 &   1.1 &    -- &    -- &  -- &   O & 0.4040 & 0.07 & 0.40 & 0.1030 & 0.22 & 1.53 \\
 J1123--4844 &     100/2276 &   93.6 &  10.6 &   1.8 &   4.8 &    -- &  -- &   O & 0.0089 & 0.13 & 0.25 & 0.0000 & 0.40 & 1.34 \\
 J1123--6102 &        5/871 &   55.0 &  10.1 &   1.5 &   5.5 &    -- &  -- &   U & 0.9846 & 0.15 & 0.32 & 0.9179 & 0.49 & 1.44 \\
 J1126--6054 &      21/2742 &   76.5 &  10.0 &   1.1 &   6.0 &    -- &  -- &   G & 0.0273 & 0.19 & 0.35 & 0.7906 & 0.67 & 1.48 \\
   J1129--53 &        6/525 &    6.5 &  31.3 &   0.2 &   9.9 &    -- &   Y &  -- & 0.3768 & 0.30 & 1.42 & 0.1604 & 0.05 & 3.67 \\
 J1133--6250 &       93/551 &  112.8 &  20.6 &   5.2 &   5.3 &    1.3483 &   Y &  -- & 0.1134 & 0.17 & 0.25 & 0.5942 & 0.49 & 1.25 \\
\hline
  \end{tabular}
   \end{centering}
}
\end{table*}

\begin{table*}
{\scriptsize
     \begin{centering}
{\small \bf{Table 1.} \emph{continued} }
 \begin{tabular}{cccccccccccccccc}
 \hline
     (1)               & (2)        & (3)          & (4)                                        & (5)                   & (6)              & (7)            & (8)              & (9)             & (10) & (11) & (12) & (13)& (14) & (15) \\
 {\bf PSR}& & {\bf S/N}&{\bf S/N} &{\bf }        &{\bf Max.}   &{\bf Min.}    & {\bf }          &{\bf Dist.}   & \\
 %\multicolumn{2}{c}{\bf best-fit} &{\bf etc.} & \\
{\bf Jname} & Npulses & {\bf (int)}&{\bf (SP)}&{\bf $\se$}&{\bf $R_{\rm j}$} &{\bf $m_{\rm j}$}& {\bf Null?}& {\bf class} & {\bf $\pln$}& {\bf $\sln$}& {\bf $\mln$}& {\bf $\pg$}&{\bf $\sg$}&{\bf $\mg$} \\
\hline
 J1136--5525 &      81/1520 &  122.5 &  17.3 &   2.7 &   6.5 &    1.7026 &  -- &   O & 0.0000 & 0.13 & 0.30 & 0.0897 & 0.41 & 1.34 \\
 J1143--5158 &        6/829 &   27.2 &  10.1 &   0.7 &   5.2 &    -- &  -- &   O & 0.0436 & 0.19 & 0.55 & 0.0345 & 0.78 & 1.91 \\
 J1146--6030 &      30/2045 &   99.5 &  18.9 &   2.0 &  10.1 &    -- &  -- &   M & 0.4683 & 0.13 & 0.32 & 0.0043 & 0.43 & 1.39 \\
 J1152--6012 &       4/1489 &   25.1 &   8.6 &   0.5 &    -- &    -- &  -- &   L & 0.9999 & 0.18 & 1.22 & 0.0000 & 1.09 & 3.52 \\
 J1157--6224 &     339/1408 &  203.0 &  25.5 &   4.1 &   7.2 &    1.1461 &   Y &  -- & 0.0000 & 0.17 & 0.20 & 0.0416 & 0.47 & 1.20 \\
 J1202--5820 &     287/1233 &  164.1 &  21.1 &   3.5 &   7.9 &    0.9868 &  -- &   L & 0.9970 & 0.14 & 0.20 & 0.0024 & 0.40 & 1.25 \\
 J1215--5328 &        1/884 &   23.4 &   7.0 &   0.7 &   4.7 &    -- &  -- &   O & 0.4205 & 0.05 & 0.65 & 0.1644 & 0.26 & 1.91 \\
 J1224--6407 &    1801/2573 &  430.1 &  38.6 &   7.0 &  14.5 &    0.7175 &   N &   O & 0.0000 & 0.09 & 0.17 & 0.0000 & 0.26 & 1.20 \\
 J1225--6035 &        5/889 &   24.9 &   7.5 &   0.6 &    -- &    -- &   Y &  -- & 0.8720 & 0.11 & 0.82 & 0.5116 & 0.59 & 2.39 \\
 J1225--6408 &       2/1327 &   73.8 &   7.1 &   1.8 &    -- &    -- &  -- &   O & 0.1341 & 0.09 & 0.15 & 0.0068 & 0.24 & 1.20 \\
 J1231--6303 &        1/407 &   48.0 &   6.2 &   2.2 &   4.6 &    -- &  -- &   O & 0.4088 & 0.14 & 0.25 & 0.0394 & 0.43 & 1.29 \\
 J1239--6832 &       11/435 &   37.9 &   8.7 &   1.4 &   5.8 &    -- &  -- &   L & 0.9460 & 0.19 & 0.25 & 0.1745 & 0.60 & 1.34 \\
 J1243--6423 &    1268/1448 & 1786.4 &  76.1 &  31.4 &   5.1 &    0.5690 &   Y & (M) & 0.0000 & 0.29 & 0.25 & 0.0000 & 0.64 & 1.15 \\
 J1252--6314 &       27/684 &   22.3 &  11.3 &   0.7 &   5.2 &    -- &  -- &   L & 0.9838 & 0.35 & 0.25 & 0.2636 & 1.09 & 1.44 \\
 J1253--5820 &     141/2190 &  168.1 &  10.6 &   2.7 &   6.0 &    1.0648 &  -- &   L & 0.7623 & 0.13 & 0.15 & 0.1696 & 0.37 & 1.20 \\
 J1255--6131 &        1/847 &    5.8 &   9.9 &   0.1 &   5.8 &    -- &   Y &  -- & 0.9963 & 0.10 & 1.97 & 0.0000 & 1.09 & 3.95 \\
 J1259--6741 &       85/840 &   80.1 &  17.0 &   2.3 &   6.0 &    -- &  -- &   L & 0.9356 & 0.21 & 0.17 & 0.0118 & 0.57 & 1.20 \\
 J1306--6617 &     138/1183 &  155.6 &  19.6 &   3.8 &    -- &    1.2613 &  -- &   M & 0.0407 & 0.15 & 0.25 & 0.0141 & 0.45 & 1.25 \\
 J1307--6318 &       11/111 &   22.8 &   9.9 &   2.1 &   4.6 &    -- &   Y &  -- & 0.9127 & 0.35 & 0.25 & 0.8660 & 1.09 & 1.01 \\
 J1312--5516 &        3/664 &   46.4 &   6.4 &   1.5 &   4.5 &    -- &  -- &   O & 0.6515 & 0.09 & 0.37 & 0.2849 & 0.33 & 1.48 \\
 J1314--6101 &        1/192 &   17.8 &   7.5 &   1.0 &    -- &    -- &  -- &   O & 0.2115 & 0.41 & 0.25 & 0.0701 & 1.09 & 1.34 \\
 J1320--5359 &       2/1985 &   76.6 &   7.7 &   1.6 &    -- &    -- &  -- &   U & 0.9936 & 0.11 & 0.40 & 0.8477 & 0.40 & 1.53 \\
 J1324--6302 &        1/196 &   13.4 &   6.7 &   0.8 &    -- &    -- &  -- &   L & 0.9973 & 0.26 & 0.72 & 0.3508 & 1.09 & 2.05 \\
 J1326--5859 &     972/1101 &  964.9 &  36.7 &  21.6 &   4.7 &    0.3409 &   Y & (L) & 0.8518 & 0.07 & 0.12 & 0.0000 & 0.20 & 1.10 \\
 J1326--6408 &       90/704 &   92.7 &  12.1 &   2.5 &   4.8 &    -- &   Y &  -- & 0.5453 & 0.19 & 0.17 & 0.4103 & 0.52 & 1.20 \\
 J1326--6700 &     682/1022 &  402.1 &  31.5 &  12.4 &   7.1 &    0.8457 &  -- &   M & 0.0000 & 0.15 & 0.27 & 0.0000 & 0.45 & 1.29 \\
 J1327--6222 &     980/1062 & 1412.3 &  76.2 &  33.0 &   5.5 &    0.4381 &   N &   M & 0.0478 & 0.15 & 0.15 & 0.0001 & 0.37 & 1.10 \\
 J1327--6301 &      89/2867 &  127.4 &  15.0 &   1.8 &    -- &    1.8150 &  -- &   O & 0.0000 & 0.15 & 0.40 & 0.0557 & 0.56 & 1.53 \\
 J1327--6400 &       9/1997 &   10.7 &   9.9 &   0.2 &    -- &    -- &  -- &   O & 0.0906 & 0.09 & 1.50 & 0.0000 & 1.09 & 3.95 \\
 J1328--4921 &        2/377 &   28.4 &   6.7 &   1.3 &    -- &    -- &  -- &   L & 0.9948 & 0.07 & 0.47 & 0.6989 & 0.26 & 1.63 \\
 J1338--6204 &       13/452 &  116.3 &   9.4 &   5.4 &   4.5 &    1.1405 &   N &   L & 0.8890 & 0.07 & 0.12 & 0.2629 & 0.20 & 1.10 \\
 J1340--6456 &      48/1470 &   53.9 &  26.9 &   1.3 &  12.6 &    -- &  -- &   O & 0.0591 & 0.11 & 0.67 & 0.0000 & 0.47 & 2.01 \\
 J1341--6023 &        2/894 &   42.6 &   7.6 &   1.0 &    -- &    -- &  -- &   L & 0.8118 & 0.11 & 0.50 & 0.5079 & 0.47 & 1.72 \\
 J1345--6115 &       15/423 &   35.4 &  10.3 &   1.4 &   4.8 &    -- &  -- &   O & 0.2355 & 0.06 & 0.32 & 0.0210 & 0.16 & 1.39 \\
 J1347--5947 &       13/898 &   43.0 &  13.9 &   1.1 &   6.8 &    -- &  -- &   O & 0.6197 & 0.18 & 0.47 & 0.0016 & 0.68 & 1.67 \\
 J1355--5153 &       36/870 &   86.5 &  12.5 &   2.0 &   6.8 &    -- &  -- &   O & 0.5448 & 0.09 & 0.27 & 0.0695 & 0.29 & 1.34 \\
   J1357--62 &     690/1236 &  396.5 &  12.0 &  11.0 &   5.1 &    0.5327 &   N &   O & 0.2966 & 0.06 & 0.15 & 0.0000 & 0.16 & 1.15 \\
 J1359--6038 &    2863/4407 &  541.4 &  18.3 &   4.5 &    -- &    0.3342 &  -- &   O & 0.0023 & 0.07 & 0.15 & 0.0003 & 0.20 & 1.15 \\
 J1401--6357 &      610/656 &  603.2 & 109.5 &  15.2 &  11.4 &    0.7267 &  -- &   M & 0.9663 & 0.18 & 0.07 & 0.0000 & 0.38 & 1.01 \\
 J1406--5806 &     111/1927 &   21.1 &  27.2 &   0.4 &   8.6 &    -- &   Y &  -- & 0.0000 & 0.25 & 1.45 & 0.0000 & 1.09 & 3.95 \\
 J1410--7404 &      21/2000 &   79.7 &  12.0 &   1.0 &   6.3 &    -- &  -- &   O & 0.0414 & 0.11 & 0.57 & 0.0007 & 0.45 & 1.86 \\
 J1413--6307 &      32/1408 &   57.5 &  22.6 &   1.1 &   9.9 &    -- &  -- &   U & 0.9496 & 0.18 & 0.57 & 0.9941 & 0.80 & 1.82 \\
 J1414--6802 &        9/112 &   37.0 &  10.1 &   2.7 &   4.8 &    -- &  -- &   O & 0.5113 & 0.17 & 0.25 & 0.0663 & 0.45 & 1.39 \\
 J1416--6037 &       6/1903 &   24.6 &   9.8 &   0.5 &    -- &    -- &  -- &   U & 0.9577 & 0.19 & 0.80 & 0.8583 & 1.09 & 2.29 \\
 J1423--6953 &     108/1676 &   35.6 &  43.6 &   0.5 &  12.4 &    -- &   Y &  -- & 0.0003 & 0.17 & 0.97 & 0.0000 & 0.99 & 2.91 \\
 J1428--5530 &      613/984 &  269.2 &  38.8 &   7.2 &  11.6 &    0.7993 &   Y & (L) & 0.9972 & 0.11 & 0.22 & 0.0000 & 0.32 & 1.25 \\
 J1430--6623 &      680/708 & 1315.8 &  78.4 &  38.3 &  11.2 &    0.4381 &   N &   L & 0.8144 & 0.15 & 0.15 & 0.0000 & 0.36 & 1.10 \\
 J1439--5501 &      1/10000 &    9.5 &   6.1 &   0.1 &    -- &    -- &  -- &   O & 0.0000 & 0.65 & 1.97 & 0.0000 & 0.10 & 2.81 \\
 J1440--6344 &       2/1228 &   67.9 &   6.5 &   1.6 &    -- &    -- &  -- &   O & 0.1214 & 0.13 & 0.30 & 0.0352 & 0.40 & 1.39 \\
 J1444--5941 &        2/204 &   16.1 &   7.1 &   0.7 &    -- &    -- &  -- &   O & 0.7104 & 0.05 & 1.17 & 0.0968 & 0.22 & 3.14 \\
 J1452--6036 &      86/3577 &   64.0 &  19.3 &   0.6 &   7.6 &    -- &  -- &   O & 0.0000 & 0.26 & 0.70 & 0.0000 & 1.09 & 2.15 \\
 J1453--6413 &    1828/3107 &  889.5 &  23.5 &   9.5 &   6.0 &    0.5077 &   Y & (L) & 0.7562 & 0.10 & 0.22 & 0.0000 & 0.28 & 1.25 \\
 J1456--6843 &    1218/2124 & 1077.9 &  41.8 &  21.5 &  15.8 &    0.8839 &  -- &   O & 0.0000 & 0.17 & 0.27 & 0.0000 & 0.43 & 1.25 \\
 J1457--5122 &       54/315 &   43.4 &  37.4 &   2.0 &   9.3 &    -- &   Y &  -- & 0.2639 & 0.38 & 0.22 & 0.0066 & 1.09 & 0.86 \\
 J1502--5653 &      96/1050 &   45.2 &  16.6 &   1.0 &   5.6 &    -- &   Y &  -- & 0.0000 & 0.11 & 0.27 & 0.0000 & 0.26 & 1.48 \\
 J1507--4352 &    1584/1952 &  380.6 &  19.0 &   6.2 &   6.2 &    0.4956 &  -- &   G & 0.0087 & 0.09 & 0.22 & 0.8712 & 0.24 & 1.25 \\
 J1507--6640 &     117/1576 &  107.8 &  12.0 &   1.6 &   6.4 &    1.2561 &  -- &   L & 0.7522 & 0.17 & 0.50 & 0.4182 & 0.64 & 1.67 \\
 J1512--5759 &      11/4367 &  189.8 &   7.9 &   2.0 &    -- &    0.9679 &  -- &   O & 0.0053 & 0.14 & 0.20 & 0.0435 & 0.40 & 1.25 \\
 J1514--4834 &     123/1224 &   99.6 &  10.9 &   2.4 &    -- &    -- &   Y &  -- & 0.3051 & 0.14 & 0.17 & 0.0051 & 0.40 & 1.20 \\
   J1514--59 &        2/533 &    5.7 &   6.6 &   0.1 &    -- &    -- &   Y &  -- & 0.1941 & 0.11 & 1.97 & 0.0000 & 1.09 & 3.95 \\
 J1522--5829 &     474/1429 &  244.5 &  12.9 &   5.9 &   5.9 &    0.7823 &  -- &   O & 0.0470 & 0.11 & 0.12 & 0.0158 & 0.30 & 1.15 \\
 J1527--3931 &       40/227 &   64.9 &  12.8 &   3.3 &   5.9 &    -- &   Y &  -- & 0.9552 & 0.13 & 0.17 & 0.1991 & 0.36 & 1.25 \\
 J1527--5552 &        6/535 &   48.7 &   7.5 &   1.6 &   4.5 &    -- &  -- &   O & 0.5225 & 0.02 & 0.20 & 0.0595 & 0.10 & 1.25 \\
 J1528--4109 &       2/1063 &   28.3 &   7.5 &   0.7 &    -- &    -- &  -- &   U & 0.9792 & 0.17 & 0.65 & 0.8310 & 0.84 & 2.01 \\
 J1530--5327 &       6/1992 &   40.7 &   7.8 &   0.8 &   4.8 &    -- &  -- &   O & 0.0364 & 0.17 & 0.60 & 0.0495 & 0.74 & 1.91 \\
 J1534--5334 &      301/410 &  194.1 &  24.4 &   6.6 &   4.8 &    0.4798 &   N &   L & 0.9280 & 0.11 & 0.05 & 0.0522 & 0.28 & 1.05 \\
 J1534--5405 &       1/1940 &   63.0 &   6.4 &   1.3 &    -- &    -- &  -- &   O & 0.0049 & 0.11 & 0.37 & 0.0000 & 0.37 & 1.48 \\
 J1535--4114 &      16/1290 &   75.1 &  12.7 &   1.8 &   8.3 &    -- &  -- &   O & 0.0772 & 0.14 & 0.25 & 0.0583 & 0.44 & 1.34 \\
 J1535--5848 &       1/1824 &   24.3 &   7.5 &   0.5 &   5.2 &    -- &  -- &   U & 0.9606 & 0.26 & 0.45 & 0.7970 & 1.09 & 1.67 \\
 J1536--5433 &        6/638 &   43.2 &   8.3 &   1.6 &   5.2 &    -- &  -- &   O & 0.0000 & 0.07 & 0.37 & 0.0000 & 0.28 & 1.44 \\
 J1539--5626 &       7/2299 &  103.9 &   6.8 &   2.0 &    -- &    1.8576 &  -- &   O & 0.1407 & 0.13 & 0.35 & 0.0096 & 0.40 & 1.44 \\
 J1539--6322 &        9/345 &   58.0 &   8.8 &   2.7 &    -- &    -- &  -- &   M & 0.8521 & 0.11 & 0.40 & 0.2512 & 0.37 & 1.48 \\
 J1542--5303 &        6/468 &   10.5 &  11.6 &   0.4 &    -- &    -- &  -- &   L & 0.9764 & 0.19 & 0.85 & 0.4618 & 1.09 & 2.48 \\
 J1544--5308 &       2/3162 &   88.2 &   9.0 &   1.2 &    -- &    -- &  -- &   O & 0.5381 & 0.13 & 0.42 & 0.1461 & 0.47 & 1.58 \\
 J1548--4927 &       86/934 &   62.0 &  19.7 &   1.5 &   7.9 &    -- &  -- &   L & 0.9758 & 0.23 & 0.45 & 0.0014 & 0.91 & 1.53 \\
 J1553--5456 &        8/522 &   33.4 &  10.6 &   1.2 &    -- &    -- &  -- &   O & 0.3615 & 0.06 & 0.32 & 0.0617 & 0.11 & 1.44 \\
 J1556--5358 &        1/561 &   21.4 &   6.8 &   0.8 &    -- &    -- &  -- &   O & 0.0368 & 0.05 & 0.60 & 0.0054 & 0.14 & 1.86 \\
 J1557--4258 &      76/1701 &  104.1 &  12.7 &   1.8 &   4.8 &    1.3912 &  -- &   O & 0.5319 & 0.14 & 0.32 & 0.3982 & 0.47 & 1.39 \\
 J1559--4438 &     561/2169 &  514.1 &  17.4 &   9.3 &   7.3 &    0.4438 &   N &   O & 0.0000 & 0.05 & 0.17 & 0.0000 & 0.14 & 1.20 \\
 J1559--5545 &       10/305 &   33.8 &  10.6 &   1.3 &   4.8 &    -- &   Y &  -- & 0.9392 & 0.27 & 0.30 & 0.5304 & 0.94 & 1.34 \\
\hline
  \end{tabular}
   \end{centering}
}
\end{table*}

\begin{table*}
{\scriptsize
     \begin{centering}
{\small \bf{Table 1.} \emph{continued} }
 \begin{tabular}{cccccccccccccccc}
 \hline
     (1)               & (2)        & (3)          & (4)                                        & (5)                   & (6)              & (7)            & (8)              & (9)             & (10) & (11) & (12) & (13)& (14) & (15) \\
 {\bf PSR}& & {\bf S/N}&{\bf S/N} &{\bf }        &{\bf Max.}   &{\bf Min.}    & {\bf }          &{\bf Dist.}   & \\
 %\multicolumn{2}{c}{\bf best-fit} &{\bf etc.} & \\
{\bf Jname} & Npulses & {\bf (int)}&{\bf (SP)}&{\bf $\se$}&{\bf $R_{\rm j}$} &{\bf $m_{\rm j}$}& {\bf Null?}& {\bf class} & {\bf $\pln$}& {\bf $\sln$}& {\bf $\mln$}& {\bf $\pg$}&{\bf $\sg$}&{\bf $\mg$} \\
\hline
 J1600--5044 &    2362/2899 &  721.5 &  28.3 &  10.5 &    -- &    0.4560 &  -- &   O & 0.0000 & 0.11 & 0.15 & 0.6171 & 0.30 & 1.15 \\
 J1602--5100 &      513/652 &  252.3 &  41.2 &   7.5 &   9.1 &    0.7918 &  -- &   M & 0.0368 & 0.13 & 0.17 & 0.0000 & 0.36 & 1.20 \\
 J1603--5657 &      13/1126 &   98.9 &   7.9 &   1.7 &    -- &    -- &  -- &   L & 0.9654 & 0.09 & 0.27 & 0.5089 & 0.29 & 1.34 \\
 J1604--4909 &     328/1725 &  165.8 &  13.7 &   3.0 &   5.6 &    0.9248 &  -- &   O & 0.4132 & 0.14 & 0.32 & 0.0000 & 0.45 & 1.39 \\
 J1605--5257 &      224/847 &  315.6 &  19.1 &  11.2 &  10.8 &    0.8175 &  -- &   L & 0.9943 & 0.11 & 0.15 & 0.0571 & 0.30 & 1.15 \\
 J1611--5847 &       1/1578 &   17.4 &   6.2 &   0.3 &    -- &    -- &  -- &   L & 0.9129 & 0.07 & 1.72 & 0.0000 & 0.02 & 3.86 \\
 J1615--5444 &       4/1549 &   33.4 &   6.9 &   0.8 &    -- &    -- &  -- &   L & 0.7997 & 0.26 & 0.35 & 0.0040 & 0.95 & 1.48 \\
 J1615--5537 &        1/686 &   15.8 &   7.2 &   0.4 &   6.5 &    -- &  -- &   O & 0.5578 & 0.15 & 0.52 & 0.1792 & 0.68 & 1.77 \\
 J1621--5039 &        1/514 &   13.5 &  11.0 &   0.5 &   6.0 &    -- &  -- &   L & 0.8536 & 0.31 & 0.45 & 0.2189 & 1.09 & 1.86 \\
 J1622--4950 &       75/130 &  263.3 &  22.5 &  22.9 &   6.4 &    0.5474 &  -- &   O & 0.0156 & 0.13 & 0.12 & 0.0035 & 0.32 & 1.15 \\
 J1624--4613 &       10/636 &   13.2 &   9.8 &   0.5 &   4.8 &    -- &   Y &  -- & 0.8338 & 0.43 & 0.45 & 0.1425 & 1.09 & 1.44 \\
 J1625--4048 &        6/230 &   27.7 &   7.6 &   1.6 &    -- &    -- &  -- &   U & 0.9853 & 0.18 & 0.45 & 0.7576 & 0.72 & 1.67 \\
 J1626--4537 &       2/1515 &   43.4 &   7.6 &   1.0 &    -- &    -- &  -- &   L & 0.8209 & 0.22 & 0.35 & 0.2921 & 0.78 & 1.48 \\
 J1632--4621 &        1/325 &   26.3 &   6.1 &   0.9 &    -- &    -- &  -- &   L & 0.9188 & 0.18 & 0.47 & 0.5621 & 0.75 & 1.67 \\
 J1633--4453 &      47/1285 &   57.2 &   9.1 &   1.4 &    -- &    -- &   Y &  -- & 0.0163 & 0.35 & 0.37 & 0.0000 & 1.09 & 1.10 \\
 J1633--5015 &     867/1590 &  340.1 &  22.1 &   7.3 &    -- &    0.5949 &  -- &   O & 0.0000 & 0.15 & 0.15 & 0.1412 & 0.38 & 1.15 \\
 J1644--4559 &    1225/1237 & 8209.0 &  43.5 & 186.1 &   7.5 &    0.1741 &   N &   O & 0.0000 & 0.07 & 0.07 & 0.0002 & 0.20 & 1.05 \\
 J1646--6831 &      195/312 &  248.1 &  64.9 &  12.5 &   9.9 &    1.0749 &   Y &  -- & 0.0000 & 0.63 & 0.40 & 0.0000 & 1.09 & 0.20 \\
   J1647--36 &      22/2672 &    9.5 &  15.1 &   0.2 &   6.5 &    -- &   Y & (L) & 0.8319 & 0.22 & 1.47 & 0.0000 & 0.13 & 3.48 \\
 J1648--3256 &       52/764 &   97.0 &   9.8 &   2.4 &   4.7 &    -- &  -- &   L & 0.9798 & 0.07 & 0.15 & 0.4164 & 0.18 & 1.20 \\
 J1649--4349 &        4/639 &   19.4 &   6.9 &   0.7 &   4.7 &    -- &   Y &  -- & 0.0034 & 0.41 & 0.55 & 0.0000 & 1.09 & 1.67 \\
 J1651--4246 &      147/664 &  385.8 &  12.8 &  14.8 &   7.1 &    0.6205 &   N &   M & 0.0002 & 0.10 & 0.12 & 0.0034 & 0.26 & 1.15 \\
 J1651--5222 &      290/878 &  157.5 &  15.0 &   4.1 &   4.9 &    0.8903 &  -- &   G & 0.0702 & 0.15 & 0.17 & 0.7991 & 0.41 & 1.20 \\
 J1651--5255 &        4/633 &   58.6 &   6.2 &   1.9 &   4.5 &    -- &  -- &   L & 0.9433 & 0.11 & 0.22 & 0.5034 & 0.36 & 1.29 \\
 J1653--3838 &     188/1848 &  110.0 &  30.6 &   2.3 &   7.6 &    1.6457 &   Y &  -- & 0.0012 & 0.15 & 0.32 & 0.0000 & 0.47 & 1.39 \\
 J1653--4249 &        2/918 &   48.5 &   7.6 &   1.3 &    -- &    -- &  -- &   O & 0.0121 & 0.07 & 0.32 & 0.0015 & 0.26 & 1.39 \\
 J1653--4854 &        1/182 &   11.8 &   7.3 &   0.6 &    -- &    -- &  -- &   L & 0.8089 & 0.30 & 0.92 & 0.7055 & 1.09 & 2.72 \\
   J1654--23 &      11/1036 &   10.9 &  13.5 &   0.3 &   9.1 &    -- &  -- &   L & 0.9931 & 0.30 & 1.02 & 0.0000 & 1.09 & 3.00 \\
 J1654--4140 &        3/427 &   24.3 &   7.0 &   1.0 &   4.6 &    -- &  -- &   M & 0.0125 & 0.14 & 0.57 & 0.0023 & 0.64 & 1.82 \\
 J1700--3312 &       62/357 &   65.8 &  16.1 &   2.7 &   8.0 &    -- &  -- &   L & 0.9846 & 0.15 & 0.25 & 0.5991 & 0.44 & 1.29 \\
 J1700--3611 &        2/371 &   28.6 &   7.9 &   1.1 &   4.5 &    -- &  -- &   O & 0.6352 & 0.13 & 0.35 & 0.1433 & 0.38 & 1.44 \\
 J1701--3130 &       7/1929 &   55.5 &   8.2 &   1.1 &    -- &    -- &  -- &   U & 0.9928 & 0.15 & 0.40 & 0.9647 & 0.56 & 1.58 \\
 J1701--3726 &      139/224 &  119.5 &  18.9 &   6.6 &   7.3 &    0.8977 &   Y &  -- & 0.0000 & 1.07 & 1.27 & 0.2732 & 0.36 & 1.86 \\
 J1701--4533 &       1/1720 &  115.1 &   6.9 &   2.8 &   4.9 &    1.8759 &  -- &   O & 0.0917 & 0.13 & 0.25 & 0.5211 & 0.36 & 1.29 \\
 J1703--3241 &      405/405 &  314.6 &  43.2 &  13.6 &   7.2 &    0.5948 &   Y & (M) & 0.0011 & 0.07 & 0.15 & 0.0000 & 0.20 & 1.15 \\
 J1703--4442 &        2/318 &   10.8 &   9.2 &   0.5 &   5.7 &    -- &  -- &   O & 0.6385 & 0.05 & 0.65 & 0.1818 & 0.11 & 2.05 \\
 J1705--1906 &     872/1867 &  242.5 &  18.8 &  13.6 &   9.6 &    0.4264 &   N &   O & 0.0000 & 0.17 & 0.15 & 0.0000 & 0.40 & 1.10 \\
         (i) &           -- &     -- &    -- &   1.8 &   8.3 &    0.7883 &   N &   L & 0.9997 & 0.13 & 0.65 & 0.0011 & 0.59 & 1.96 \\
         (m) &           -- &     -- &    -- &  14.2 &   9.6 &    0.4264 &   N &   O & 0.0000 & 0.18 & -0.00 & 0.0000 & 0.37 & 0.96 \\
 J1705--3423 &     187/2238 &  188.1 &  10.4 &   3.8 &   5.3 &    1.3555 &  -- &   O & 0.0029 & 0.15 & 0.12 & 0.1837 & 0.40 & 1.15 \\
 J1705--3950 &       7/1746 &   35.6 &   9.8 &   0.7 &   4.8 &    -- &  -- &   L & 0.8104 & 0.26 & 0.50 & 0.0063 & 1.09 & 1.72 \\
 J1706--6118 &     119/1534 &   52.1 &  34.5 &   0.8 &   8.7 &    -- &  -- &   O & 0.6280 & 0.17 & 0.72 & 0.0000 & 0.75 & 2.24 \\
 J1707--4053 &       16/960 &  124.9 &   9.2 &   3.5 &    -- &    1.3939 &  -- &   O & 0.6191 & 0.10 & 0.25 & 0.3935 & 0.29 & 1.29 \\
   J1707--44 &         3/98 &   13.5 &   8.2 &   1.1 &   5.3 &    -- &  -- &   L & 0.9345 & 0.09 & 0.70 & 0.7019 & 0.48 & 2.05 \\
 J1707--4729 &      68/2084 &   60.8 &  14.2 &   1.3 &    -- &    -- &   Y &  -- & 0.0000 & 0.31 & 0.37 & 0.1334 & 1.09 & 1.34 \\
 J1708--3426 &       19/802 &   62.2 &  10.9 &   1.9 &   5.6 &    -- &  -- &   O & 0.3210 & 0.15 & 0.22 & 0.0445 & 0.45 & 1.29 \\
 J1709--1640 &      627/857 &  386.2 &  37.2 &  10.1 &  12.9 &    0.7796 &   Y &  -- & 0.0000 & 0.21 & 0.25 & 0.0006 & 0.60 & 1.20 \\
 J1709--4429 &      31/5438 &  198.3 &   7.5 &   2.3 &   5.9 &    1.3016 &  -- &   O & 0.0001 & 0.13 & 0.22 & 0.0001 & 0.38 & 1.29 \\
 J1711--5350 &        7/614 &   53.2 &   7.9 &   1.6 &    -- &    -- &  -- &   L & 0.9692 & 0.15 & 0.35 & 0.6167 & 0.53 & 1.44 \\
 J1715--4034 &       14/270 &   49.5 &  11.3 &   2.7 &    -- &    -- &  -- &   O & 0.0401 & 0.18 & 0.25 & 0.0004 & 0.53 & 1.34 \\
 J1717--3425 &        2/848 &   94.3 &   7.0 &   2.8 &    -- &    -- &  -- &   O & 0.2274 & 0.02 & 0.20 & 0.0019 & 0.09 & 1.20 \\
 J1717--4043 &       7/1416 &   17.4 &   8.2 &   0.5 &    -- &    -- &  -- &   L & 0.8154 & 0.23 & 0.87 & 0.0000 & 1.09 & 2.58 \\
 J1718--3825 &       1/7607 &   29.5 &   7.7 &   0.3 &    -- &    -- &  -- &   O & 0.0000 & 0.30 & 1.27 & 0.0000 & 1.09 & 3.76 \\
 J1720--2933 &        1/885 &   31.5 &   6.4 &   0.9 &   5.0 &    -- &  -- &   L & 0.8275 & 0.09 & 0.42 & 0.4992 & 0.26 & 1.53 \\
 J1721--3532 &      15/1987 &  220.1 &   7.9 &   4.6 &   4.9 &    1.4397 &  -- &   O & 0.0000 & 0.15 & 0.22 & 0.0001 & 0.44 & 1.25 \\
 J1722--3207 &     652/1170 &  265.2 &  50.0 &   6.8 &  14.9 &    0.7151 &   N &   O & 0.1234 & 0.09 & 0.20 & 0.0000 & 0.25 & 1.20 \\
 J1722--3632 &       1/1413 &   37.9 &   6.5 &   1.0 &    -- &    -- &  -- &   O & 0.4579 & 0.21 & 0.47 & 0.3012 & 0.84 & 1.67 \\
 J1722--3712 &     241/2382 &  214.7 &  10.3 &   3.3 &    -- &    0.8111 &  -- &   O & 0.0361 & 0.05 & 0.65 & 0.1161 & 0.22 & 1.96 \\
 J1723--3659 &       1/2740 &   38.5 &  14.5 &   0.7 &    -- &    -- &  -- &   G & 0.0745 & 0.25 & 0.42 & 0.7637 & 1.01 & 1.63 \\
 J1725--4043 &        8/368 &   18.4 &  15.5 &   0.8 &   8.1 &    -- &   Y &  -- & 0.9392 & 0.34 & 0.60 & 0.0040 & 1.09 & 1.77 \\
 J1727--2739 &      127/434 &   74.6 &  28.4 &   3.8 &   7.3 &    -- &   Y &  -- & 0.0000 & 0.57 & 0.52 & 0.0026 & 1.09 & 0.91 \\
 J1730--3350 &      22/4057 &   83.4 &  12.6 &   1.0 &   4.8 &    -- &  -- &   O & 0.0000 & 0.26 & 0.57 & 0.0103 & 1.09 & 1.72 \\
 J1731--4744 &      593/674 & 1167.8 & 106.8 &  33.1 &  13.9 &    0.5113 &   N &   L & 0.9826 & 0.14 & 0.15 & 0.0055 & 0.36 & 1.15 \\
 J1732--4156 &       1/1736 &    8.7 &   8.4 &   0.2 &   6.0 &    -- &  -- &   O & 0.0762 & 0.46 & 0.90 & 0.0000 & 1.09 & 3.95 \\
 J1733--2228 &       19/645 &  136.4 &   7.8 &   5.3 &   4.7 &    1.0549 &  -- &   L & 0.8932 & 0.09 & 0.15 & 0.1185 & 0.24 & 1.20 \\
 J1733--3716 &     146/1663 &   53.8 &  32.9 &   1.4 &   9.9 &    -- &  -- &   O & 0.0002 & 0.23 & 0.32 & 0.0000 & 0.74 & 1.39 \\
 J1735--0724 &      13/1318 &   96.8 &   7.0 &   2.3 &    -- &    -- &  -- &   U & 0.9983 & 0.09 & 0.45 & 0.9913 & 0.32 & 1.58 \\
 J1736--2457 &       25/211 &   40.3 &  11.4 &   2.2 &   4.6 &    -- &   Y &  -- & 0.9564 & 0.21 & 0.22 & 0.2750 & 0.56 & 1.25 \\
 J1737--3555 &       3/1417 &   30.8 &   7.0 &   0.7 &    -- &    -- &  -- &   O & 0.6267 & 0.22 & 0.52 & 0.0172 & 0.92 & 1.77 \\
 J1738--2330 &        7/277 &   13.1 &   8.1 &   0.7 &   4.8 &    -- &   Y &  -- & 0.9480 & 0.33 & 0.40 & 0.5859 & 1.06 & 1.82 \\
 J1738--3211 &      162/732 &  105.7 &  27.9 &   2.9 &  14.6 &    1.3554 &   Y &  -- & 0.0496 & 0.17 & 0.32 & 0.1170 & 0.52 & 1.39 \\
 J1739--2903 &     104/1739 &   82.2 &  20.8 &   2.1 &  11.6 &    -- &  -- &   O & 0.0024 & 0.13 & 0.25 & 0.0000 & 0.38 & 1.29 \\
         (i) &           -- &     -- &    -- &   0.9 &    -- &    -- &   N &   L & 0.9197 & 0.14 & 0.55 & 0.5745 & 0.55 & 1.82 \\
         (m) &           -- &     -- &    -- &   1.9 &  11.6 &    -- &   N &   O & 0.3612 & 0.17 & 0.45 & 0.0000 & 0.57 & 1.58 \\
 J1739--3023 &       1/4870 &   13.8 &   6.1 &   0.1 &    -- &    -- &  -- &   O & 0.0191 & 0.25 & 1.35 & 0.0000 & 0.13 & 3.95 \\
 J1740--3015 &      480/919 &  179.6 &  19.8 &   3.5 &   6.8 &    0.6859 &  -- &   O & 0.0574 & 0.10 & 0.22 & 0.0002 & 0.28 & 1.29 \\
 J1741--0840 &      154/272 &  121.7 &  27.1 &   6.3 &   6.6 &    1.0280 &   Y &  -- & 0.0000 & 0.81 & 0.82 & 0.0000 & 1.09 & 0.20 \\
 J1741--2019 &       29/135 &   31.6 &  19.8 &   2.0 &   6.3 &    -- &  -- &   O & 0.7164 & 0.21 & 0.20 & 0.4152 & 0.59 & 1.25 \\
 J1741--3016 &        9/297 &   41.1 &   9.3 &   2.1 &   4.6 &    -- &   Y &  -- & 0.9920 & 0.14 & 0.32 & 0.5881 & 0.49 & 1.44 \\
 J1741--3927 &     375/1015 &  200.4 &  17.5 &   5.3 &   4.7 &    0.6947 &   N &   M & 0.1270 & 0.11 & 0.15 & 0.0000 & 0.30 & 1.15 \\
\hline
  \end{tabular}
   \end{centering}
}
\end{table*}

\begin{table*}
{\scriptsize
     \begin{centering}
{\small \bf{Table 1.} \emph{continued} }
 \begin{tabular}{cccccccccccccccc}
 \hline
     (1)               & (2)        & (3)          & (4)                                        & (5)                   & (6)              & (7)            & (8)              & (9)             & (10) & (11) & (12) & (13)& (14) & (15) \\
 {\bf PSR}& & {\bf S/N}&{\bf S/N} &{\bf }        &{\bf Max.}   &{\bf Min.}    & {\bf }          &{\bf Dist.}   & \\
 %\multicolumn{2}{c}{\bf best-fit} &{\bf etc.} & \\
{\bf Jname} & Npulses & {\bf (int)}&{\bf (SP)}&{\bf $\se$}&{\bf $R_{\rm j}$} &{\bf $m_{\rm j}$}& {\bf Null?}& {\bf class} & {\bf $\pln$}& {\bf $\sln$}& {\bf $\mln$}& {\bf $\pg$}&{\bf $\sg$}&{\bf $\mg$} \\
\hline
 J1742--4616 &      10/1357 &   56.2 &   7.5 &   1.5 &   5.3 &    -- &   Y &  -- & 0.0000 & 0.31 & 0.35 & 0.0790 & 1.09 & 1.25 \\
 J1743--3150 &       86/233 &   85.5 &  20.9 &   4.1 &   6.3 &    -- &  -- &   L & 0.9926 & 0.19 & 0.10 & 0.5792 & 0.45 & 1.10 \\
 J1744--1134 &    16/138924 &   22.4 &  11.0 &   0.0 &    -- &    -- &  -- &   O & 0.0000 & 0.38 & 1.97 & 0.0000 & 0.05 & 2.34 \\
 J1744--1610 &        3/320 &   18.5 &   8.1 &   0.8 &   4.8 &    -- &  -- &   L & 0.9892 & 0.22 & 0.45 & 0.6995 & 0.91 & 1.63 \\
 J1744--3130 &        4/528 &   18.2 &   8.0 &   0.6 &   4.8 &    -- &  -- &   O & 0.0181 & 0.09 & 0.32 & 0.0005 & 0.22 & 1.48 \\
 J1745--3040 &     738/1529 &  408.3 &  79.5 &   7.7 &  11.0 &    1.3502 &   Y &  -- & 0.0000 & 0.34 & -0.13 & 0.0000 & 0.41 & 0.72 \\
 J1749--5605 &       22/418 &   47.7 &  16.1 &   1.6 &   6.4 &    -- &  -- &   O & 0.7369 & 0.17 & 0.20 & 0.1151 & 0.49 & 1.29 \\
 J1750--3157 &       42/610 &   41.1 &  13.3 &   1.5 &   5.9 &    -- &   Y &  -- & 0.0058 & 0.27 & 0.50 & 0.0935 & 1.09 & 1.48 \\
 J1751--3323 &       2/1014 &   47.7 &   8.4 &   1.3 &    -- &    -- &  -- &   O & 0.7010 & 0.15 & 0.30 & 0.3216 & 0.51 & 1.39 \\
 J1751--4657 &      179/739 &  133.6 &  31.9 &   3.6 &  13.3 &    0.9885 &  -- &   L & 0.9992 & 0.11 & 0.45 & 0.2437 & 0.43 & 1.58 \\
 J1752--2806 &      938/995 & 1304.1 &  66.7 &  26.6 &   8.3 &    0.4962 &   N &   O & 0.0531 & 0.21 & 0.12 & 0.0000 & 0.48 & 1.10 \\
   J1753--38 &       14/848 &   20.1 &  16.0 &   0.4 &   8.8 &    -- &  -- &   O & 0.0820 & 0.38 & 0.22 & 0.0000 & 1.09 & 1.72 \\
 J1754--3510 &      31/1421 &   41.2 &  11.0 &   0.8 &   6.1 &    -- &  -- &   O & 0.3901 & 0.18 & 0.57 & 0.0259 & 0.80 & 1.86 \\
 J1755--2521 &        1/465 &   16.7 &   9.2 &   0.6 &   4.4 &    -- &  -- &   L & 0.8487 & 0.10 & 0.77 & 0.3250 & 0.53 & 2.29 \\
 J1756--2225 &       9/1387 &   11.4 &  10.9 &   0.3 &    -- &    -- &  -- &   L & 0.9745 & 0.19 & 1.27 & 0.0001 & 1.09 & 3.81 \\
 J1756--2435 &        1/827 &   70.6 &   6.5 &   2.2 &    -- &    -- &  -- &   O & 0.0273 & 0.14 & 0.17 & 0.0176 & 0.38 & 1.20 \\
 J1757--2223 &      40/3042 &   17.5 &  19.7 &   0.2 &   8.6 &    -- &   Y &  -- & 0.0284 & 0.27 & 1.00 & 0.0000 & 1.09 & 3.95 \\
 J1757--2421 &     168/2380 &  197.4 &  17.4 &   3.6 &   9.1 &    1.3102 &  -- &   O & 0.7076 & 0.10 & 0.27 & 0.0242 & 0.32 & 1.29 \\
 J1758--2540 &       12/266 &   32.6 &   8.7 &   1.9 &    -- &    -- &   Y &  -- & 0.0028 & 0.66 & 0.82 & 0.0017 & 1.09 & 1.10 \\
 J1758--2846 &        1/725 &   13.8 &   6.1 &   0.4 &    -- &    -- &  -- &   O & 0.1359 & 0.05 & 0.90 & 0.0358 & 0.18 & 2.53 \\
 J1759--1956 &       19/194 &   43.1 &   9.7 &   2.0 &   5.9 &    -- &  -- &   O & 0.6426 & 0.21 & 0.37 & 0.0449 & 0.68 & 1.48 \\
 J1759--2205 &     181/1222 &  133.1 &  15.1 &   2.1 &   5.0 &    0.7153 &  -- &   O & 0.0001 & 0.09 & 0.27 & 0.0000 & 0.25 & 1.29 \\
 J1759--3107 &       13/511 &   48.9 &   8.9 &   1.6 &    -- &    -- &  -- &   U & 0.9915 & 0.14 & 0.32 & 0.8948 & 0.47 & 1.44 \\
 J1801--2920 &       42/514 &   48.5 &  18.3 &   1.9 &   7.6 &    -- &  -- &   L & 0.9067 & 0.18 & 0.45 & 0.0076 & 0.67 & 1.53 \\
 J1803--1857 &        7/191 &   29.2 &  12.3 &   1.3 &   4.3 &    -- &  -- &   O & 0.7172 & 0.11 & 0.37 & 0.1624 & 0.41 & 1.48 \\
 J1803--2137 &      67/4166 &  158.6 &  11.7 &   1.9 &   5.3 &    1.7331 &  -- &   O & 0.0000 & 0.21 & 0.40 & 0.0000 & 0.69 & 1.44 \\
 J1805--1504 &        1/357 &   81.2 &   6.4 &   4.3 &    -- &    -- &  -- &   U & 0.8400 & 0.15 & 0.30 & 0.9497 & 0.47 & 1.34 \\
 J1806--1154 &       1/1071 &   75.8 &   6.1 &   2.2 &    -- &    -- &  -- &   L & 0.9721 & 0.13 & 0.22 & 0.2183 & 0.36 & 1.29 \\
 J1807--0847 &     824/3403 &  390.2 &  10.9 &   5.4 &   5.9 &    0.4645 &  -- &   G & 0.0105 & 0.07 & 0.20 & 0.7922 & 0.20 & 1.25 \\
 J1807--2715 &       11/676 &   56.2 &  11.4 &   1.7 &   4.7 &    -- &  -- &   U & 0.9929 & 0.11 & 0.30 & 0.8543 & 0.37 & 1.39 \\
 J1808--0813 &        2/629 &   53.0 &   8.4 &   1.8 &    -- &    -- &   Y &  -- & 0.0425 & 0.18 & 0.30 & 0.1071 & 0.60 & 1.39 \\
 J1808--2057 &       41/603 &  108.3 &  11.9 &   3.2 &   4.5 &    1.1464 &  -- &   L & 0.9894 & 0.15 & 0.20 & 0.2860 & 0.43 & 1.25 \\
 J1808--3249 &       3/1544 &   34.8 &   6.7 &   0.8 &   5.1 &    -- &  -- &   O & 0.0001 & 0.11 & 0.42 & 0.0000 & 0.45 & 1.58 \\
 J1809--2109 &       35/795 &   41.0 &  12.1 &   1.0 &   6.6 &    -- &   Y &  -- & 0.0181 & 0.38 & 0.52 & 0.0000 & 1.09 & 1.48 \\
 J1814--0618 &        1/109 &   17.6 &   6.9 &   1.4 &    -- &    -- &  -- &   U & 0.8606 & 0.39 & 0.27 & 0.9719 & 1.09 & 1.10 \\
 J1814--1649 &        1/593 &   41.7 &   6.9 &   1.5 &    -- &    -- &  -- &   L & 0.9840 & 0.14 & 0.30 & 0.5557 & 0.45 & 1.39 \\
 J1815--1910 &        1/449 &    9.2 &   6.3 &   0.3 &    -- &    -- &  -- &   O & 0.4623 & 0.38 & 0.60 & 0.0241 & 1.09 & 2.24 \\
 J1816--1729 &        3/713 &   59.4 &   7.3 &   1.7 &    -- &    -- &  -- &   L & 0.8081 & 0.09 & 0.35 & 0.1825 & 0.29 & 1.44 \\
 J1817--3618 &     279/1453 &  114.4 &  29.1 &   2.3 &  13.1 &    1.5808 &   Y &  -- & 0.0000 & 0.33 & 0.22 & 0.0031 & 0.92 & 1.10 \\
 J1817--3837 &      17/1465 &   85.4 &   8.3 &   1.6 &   4.7 &    -- &  -- &   L & 0.9383 & 0.11 & 0.40 & 0.4887 & 0.37 & 1.53 \\
 J1819--1458 &        7/132 &    9.7 &  16.8 &   0.5 &   8.1 &    -- &   Y & (L) & 0.8941 & 0.06 & 0.40 & 0.3709 & 0.18 & 1.53 \\
 J1820--0427 &      913/926 &  595.9 &  43.1 &  14.7 &   7.0 &    0.3359 &   N &   L & 0.8796 & 0.09 & 0.12 & 0.0151 & 0.22 & 1.15 \\
 J1820--0509 &      22/1661 &   17.8 &   9.8 &   0.4 &   4.9 &    -- &   Y &  -- & 0.0477 & 0.26 & 1.15 & 0.0000 & 1.09 & 3.33 \\
 J1820--1346 &       29/592 &   76.5 &  10.1 &   2.8 &    -- &    -- &  -- &   L & 0.9107 & 0.17 & 0.25 & 0.1762 & 0.48 & 1.29 \\
 J1821--1432 &        1/273 &   12.9 &   7.9 &   0.7 &    -- &    -- &  -- &   O & 0.0125 & 0.06 & 0.47 & 0.0001 & 0.07 & 1.67 \\
 J1822--2256 &      175/279 &  103.3 &  17.1 &   5.1 &   6.5 &    1.0380 &   Y & (U) & 1.0000 & 0.10 & 0.20 & 0.7840 & 0.29 & 1.20 \\
 J1823--1126 &        9/292 &   20.5 &  20.6 &   0.9 &   7.0 &    -- &   Y & (L) & 0.9112 & 0.30 & 0.12 & 0.0420 & 0.40 & 1.44 \\
 J1824--1945 &    2009/2962 &  454.2 &  24.9 &   5.4 &   6.2 &    0.4280 &   N &   O & 0.1838 & 0.09 & 0.20 & 0.0000 & 0.24 & 1.20 \\
 J1824--2233 &        2/456 &   21.2 &   8.2 &   0.7 &    -- &    -- &  -- &   O & 0.6926 & 0.49 & 0.17 & 0.0162 & 1.09 & 1.20 \\
 J1824--2328 &        1/329 &   25.1 &   6.8 &   1.1 &    -- &    -- &  -- &   L & 0.8165 & 0.22 & 0.40 & 0.3816 & 0.80 & 1.53 \\
 J1825--0935 &      566/731 &  358.9 &  69.5 &   8.1 &  14.5 &    0.7238 &  -- &   O & 0.4060 & 0.18 & 0.20 & 0.0000 & 0.47 & 1.20 \\
         (i) &           -- &     -- &    -- &   1.0 &    -- &    1.9292 &   N &   L & 0.9755 & 0.18 & 0.52 & 0.1689 & 0.68 & 1.77 \\
         (m) &           -- &     -- &    -- &   8.3 &  14.5 &    0.7238 &   N &   O & 0.0166 & 0.15 & 0.17 & 0.0000 & 0.37 & 1.15 \\
 J1825--1446 &     236/2016 &  110.0 &  50.4 &   2.2 &   6.5 &    2.3894 &   Y &  -- & 0.0000 & 0.17 & 0.27 & 0.0000 & 0.41 & 1.29 \\
 J1826--1131 &        1/267 &   60.3 &   7.2 &   3.3 &    -- &    -- &  -- &   L & 0.9887 & 0.02 & 0.30 & 0.4725 & 0.07 & 1.34 \\
 J1827--0750 &      45/2070 &   48.7 &  17.9 &   1.0 &   5.5 &    -- &   Y &  -- & 0.0000 & 0.17 & 1.20 & 0.0000 & 1.09 & 3.14 \\
 J1829--0734 &       5/1750 &   19.6 &   8.0 &   0.4 &   5.1 &    -- &  -- &   O & 0.0046 & 0.25 & 0.90 & 0.0000 & 1.09 & 2.72 \\
 J1829--1751 &     615/1820 &  301.7 &  17.2 &   6.3 &   9.0 &    0.7746 &  -- &   L & 0.9018 & 0.09 & 0.20 & 0.0016 & 0.25 & 1.25 \\
 J1830--1059 &      10/1370 &   48.3 &   9.3 &   0.9 &    -- &    -- &  -- &   O & 0.7474 & 0.18 & 0.35 & 0.5392 & 0.67 & 1.48 \\
 J1830--1135 &        16/78 &   31.0 &  12.4 &   2.6 &   4.6 &    -- &   Y &  -- & 0.2407 & 0.25 & 0.27 & 0.2606 & 0.72 & 1.24 \\
 J1831--1223 &       15/193 &   29.2 &  10.1 &   1.8 &   5.0 &    -- &   Y &  -- & 0.9205 & 0.22 & 0.42 & 0.2856 & 0.90 & 1.48 \\
 J1831--1329 &        3/253 &   29.2 &   6.7 &   1.5 &    -- &    -- &  -- &   U & 0.9920 & 0.18 & 0.37 & 0.7959 & 0.60 & 1.48 \\
 J1832--0827 &      111/875 &  113.6 &  15.0 &   2.7 &   6.3 &    0.8993 &  -- &   L & 0.8204 & 0.13 & 0.15 & 0.0827 & 0.34 & 1.20 \\
 J1833--0338 &      296/818 &  153.1 &  13.9 &   3.4 &   9.0 &    0.6031 &  -- &   O & 0.2235 & 0.10 & 0.15 & 0.0003 & 0.26 & 1.15 \\
 J1833--0827 &     288/6566 &  110.6 &  19.4 &   0.8 &    -- &    1.8933 &  -- &   O & 0.0000 & 0.27 & 0.75 & 0.0000 & 1.09 & 2.05 \\
 J1833--1055 &        2/869 &   15.3 &   6.6 &   0.5 &    -- &    -- &  -- &   O & 0.1725 & 0.35 & 0.75 & 0.0000 & 1.09 & 2.24 \\
 J1834--0426 &      12/1951 &  277.3 &   9.1 &   6.3 &   6.3 &    1.5350 &  -- &   O & 0.0000 & 0.09 & 0.20 & 0.1849 & 0.25 & 1.20 \\
 J1835--1020 &        1/609 &   40.9 &   6.7 &   1.4 &    -- &    -- &  -- &   U & 1.0000 & 0.13 & 0.32 & 0.9650 & 0.41 & 1.39 \\
 J1835--1106 &       8/3393 &   83.3 &   8.4 &   0.9 &   5.1 &    -- &  -- &   L & 0.9600 & 0.11 & 0.85 & 0.3537 & 0.65 & 2.39 \\
 J1836--0436 &       3/1582 &   75.1 &   7.2 &   1.6 &    -- &    -- &  -- &   O & 0.0684 & 0.09 & 0.37 & 0.0070 & 0.29 & 1.48 \\
 J1836--1008 &      255/995 &  172.2 &  10.9 &   4.0 &    -- &    0.5883 &  -- &   O & 0.5187 & 0.05 & 0.17 & 0.0001 & 0.16 & 1.20 \\
 J1837--0653 &       79/290 &   72.1 &  24.4 &   3.5 &   6.1 &    -- &   Y &  -- & 0.0000 & 0.81 & 1.30 & 0.0000 & 1.09 & 0.91 \\
 J1837--1243 &        2/285 &    6.4 &   8.6 &   0.2 &    -- &    -- &   Y &  -- & 0.9562 & 0.33 & 1.20 & 0.0432 & 1.09 & 3.71 \\
 J1839--1238 &        1/291 &   25.6 &   8.1 &   1.2 &   5.2 &    -- &  -- &   O & 0.1427 & 0.04 & 0.40 & 0.0089 & 0.13 & 1.48 \\
 J1840--0809 &       64/568 &   75.3 &  12.3 &   2.5 &   4.9 &    -- &  -- &   L & 0.8623 & 0.11 & 0.47 & 0.5718 & 0.44 & 1.63 \\
 J1840--0815 &       55/477 &   82.2 &  18.9 &   2.7 &   6.7 &    -- &  -- &   L & 0.8874 & 0.13 & 0.25 & 0.2749 & 0.37 & 1.29 \\
 J1840--0840 &       42/101 &   72.3 &  33.2 &   6.6 &   6.3 &    -- &   Y &  -- & 0.4169 & 0.49 & 0.50 & 0.4267 & 1.09 & 0.91 \\
 J1840--1417 &         9/85 &   28.8 &  68.9 &   1.3 &   7.3 &    -- &   Y &  -- & 0.9129 & 0.05 & 0.47 & 0.4385 & 0.29 & 1.63 \\
 J1841--0157 &        9/846 &   47.4 &   8.2 &   1.5 &   5.1 &    -- &  -- &   O & 0.5375 & 0.17 & 0.35 & 0.0186 & 0.53 & 1.44 \\
 J1841--0310 &        4/334 &    5.7 &   9.7 &   0.3 &   5.5 &    -- &   Y &  -- & 0.9858 & 0.10 & 1.25 & 0.5221 & 0.53 & 3.95 \\
\hline
  \end{tabular}
   \end{centering}
}
\end{table*}

\begin{table*}
{\scriptsize
     \begin{centering}
{\small \bf{Table 1.} \emph{continued} }
 \begin{tabular}{cccccccccccccccc}
 \hline
     (1)               & (2)        & (3)          & (4)                                        & (5)                   & (6)              & (7)            & (8)              & (9)             & (10) & (11) & (12) & (13)& (14) & (15) \\
 {\bf PSR}& & {\bf S/N}&{\bf S/N} &{\bf }        &{\bf Max.}   &{\bf Min.}    & {\bf }          &{\bf Dist.}   & \\
 %\multicolumn{2}{c}{\bf best-fit} &{\bf etc.} & \\
{\bf Jname} & Npulses & {\bf (int)}&{\bf (SP)}&{\bf $\se$}&{\bf $R_{\rm j}$} &{\bf $m_{\rm j}$}& {\bf Null?}& {\bf class} & {\bf $\pln$}& {\bf $\sln$}& {\bf $\mln$}& {\bf $\pg$}&{\bf $\sg$}&{\bf $\mg$} \\
\hline
 J1841--0425 &      32/3016 &  128.2 &   8.9 &   1.9 &    -- &    1.7033 &  -- &   O & 0.5864 & 0.10 & 0.32 & 0.0001 & 0.33 & 1.39 \\
 J1842--0359 &       92/305 &  110.5 &  20.4 &   7.0 &   6.2 &    1.1869 &  -- &   L & 0.9958 & 0.17 & 0.20 & 0.1326 & 0.44 & 1.20 \\
 J1843--0459 &        1/734 &   45.8 &   6.3 &   1.6 &   4.8 &    -- &  -- &   G & 0.0409 & 0.26 & 0.30 & 0.8523 & 0.95 & 1.29 \\
 J1844--0433 &       53/570 &   76.6 &  15.1 &   2.3 &   5.6 &    -- &  -- &   O & 0.1971 & 0.15 & 0.30 & 0.2807 & 0.49 & 1.34 \\
 J1845--0434 &      15/1149 &  113.5 &   7.6 &   2.9 &    -- &    1.2484 &  -- &   L & 0.8873 & 0.13 & 0.12 & 0.1470 & 0.32 & 1.15 \\
J1846--07492 &        1/655 &   28.2 &   6.1 &   0.9 &    -- &    -- &  -- &   L & 0.9840 & 0.11 & 0.57 & 0.5607 & 0.44 & 1.82 \\
 J1847--0402 &       59/927 &  124.3 &  10.6 &   3.4 &   5.1 &    1.1194 &  -- &   O & 0.0243 & 0.09 & 0.15 & 0.0002 & 0.25 & 1.20 \\
 J1847--0605 &        4/686 &   13.7 &   7.3 &   0.5 &   5.0 &    -- &  -- &   L & 0.9282 & 0.13 & 0.72 & 0.2881 & 0.59 & 2.15 \\
 J1848--1150 &        3/417 &   15.7 &   8.5 &   0.6 &    -- &    -- &  -- &   O & 0.3449 & 0.35 & 0.40 & 0.0009 & 1.09 & 1.63 \\
 J1848--1952 &       58/122 &  137.2 &  91.5 &   8.0 &   7.4 &    1.0877 &   Y &  -- & 0.0084 & 0.42 & -0.13 & 0.0175 & 0.87 & 0.53 \\
 J1852--0635 &     150/1067 &   87.7 &  21.8 &   2.7 &   9.1 &    -- &   Y &  -- & 0.4344 & 0.23 & 0.10 & 0.0000 & 0.51 & 1.10 \\
 J1854--1421 &      149/476 &  116.8 &  22.9 &   4.2 &   8.7 &    0.8327 &  -- &   O & 0.1650 & 0.15 & 0.12 & 0.0351 & 0.40 & 1.15 \\
 J1857--1027 &       73/145 &   92.5 &  41.0 &   6.4 &   7.3 &    -- &   Y &  -- & 0.1064 & 0.34 & 0.15 & 0.1667 & 0.78 & 1.05 \\
 J1900--2600 &      548/919 &  385.8 &  21.9 &  12.6 &   8.4 &    0.7765 &  -- &   M & 0.0003 & 0.09 & 0.27 & 0.0000 & 0.29 & 1.29 \\
 J1901--0906 &      219/313 &  156.5 &  58.0 &   6.8 &   7.1 &    0.8386 &  -- &   U & 1.0000 & 0.11 & 0.22 & 0.9777 & 0.30 & 1.25 \\
 J1901--1740 &        5/285 &   17.3 &   7.7 &   0.8 &    -- &    -- &  -- &   L & 0.9654 & 0.39 & 0.15 & 0.5161 & 1.09 & 1.15 \\
 J1903--0632 &       8/1304 &   47.1 &   9.0 &   1.0 &    -- &    -- &  -- &   L & 0.7572 & 0.15 & 0.47 & 0.2843 & 0.61 & 1.67 \\

\hline
\end{tabular}
\end{centering}
}
\end{table*}

\section{Single-Pulse Energy Distributions}\label{sec:edists}
Here we describe the results of our energy distribution shape-fit tests, with the goal of characterising the field statistics of the radio-generating pulsar plasma. Section \ref{sec:eclass} organises the pulsars into categories defined by their energy distribution shape. Section \ref{sec:estats} interprets these class divisions in terms of underlying pulsar energy statistics, taking into account our data's noise properties and other caveats of the fitting analysis. That section also reviews the typical distribution parameters defining the best-fit pulsar shapes. Finally, Section \ref{sec:bimodal} explores the cause of the distinct, multiple, non-zero energy peaks exhibited by some pulsars in our sample.

\subsection{Classification of energy distributions}\label{sec:eclass}
Table \ref{table:stats} reports our classifications (described below) for each pulsar, along with the best-fit parameters and goodness-of-fit probability for the Gaussian and log-normal fits.

During visual inspection of the energy distributions, we noted multiple non-null peaks in the distributions of some pulsars. As with nulling pulsars, these are not unimodal and thus their best-fit distribution statistics are not reliable. We provide the fit results in Table\,\ref{table:stats} for nulling and multi-peaked pulsars only for the sake of the \S\ref{sec:estats} discussion.
Several nulling pulsars had a sufficiently bright non-null state to have a recognisably distinct distribution from the null pulses. These are identifiable in Table \ref{table:stats} as pulsars with a distribution class (column 9) reported in parentheses. An example of such a case is shown in Figure \ref{fig:nulldist}. These objects are included in our statistical analyses.
Multi-peaked energy distributions were defined as any distribution with either more than two points deviating more than one standard deviation, or one point deviating more than two standard deviations from a smooth single-peaked distribution. These were identified by inspection of the pulse energy plots.

We divided the pulsars into five energy distribution classifications (the percentage of constituent pulsars is given for each category; these percentages are calculated based on the 255 non-nulling, and classifiable-nulling, pulsars):
\begin{itemize} % NOTE: 315 pulsars total, but 255 classifiable.
	\item \bi{Log-normal (L; 33\%)} Distribution appeared unimodal and the best-fit results obeyed $\pln\geq0.75$, and $\pg<0.75$
	% underestimated a right-ended tail, and the best log-normal did better than gaussian...?.
	\item \bi{Gaussian (G; 3\%)} Distribution appeared unimodal and the best-fit results obeyed $\pln<0.75$, and $\pg\geq0.75$
	\item \bi{Unimodal (U; 9\%)} $\pln$ and $\pg$ were $\geq$0.75
	\item \bi{Multi-peaked (M; 7\%)} As described above, two or more peaks were discernible at energy levels above the noise % , indicating that there are several energy states apparent, by visual inspection of the PED. M* = multipeaked nuller. MB = bimodal.
	\item \bi{Other (O; 48\%)} Pulsars with $\pln,\pg<0.75$\end{itemize}
% these may be nulling RRATs but we don't have power (i.e. enough N or sensitivity) to distinguish log-normal from null. Note we can still use these in the paper for mod parameters like R-param, at least to get a lower limit!). Those with integrated energy we might say non-nulling, but for those with no integrated energy we can say these are deep nulling pulsars; for instance, my ``intermittent,'' J0941--39-like ones should fall into this category. some had too few pulses in distribution to tell anything (no bins of significant enough expected value to make a sensible $\chi^2$ estimate)
In cases where we had multiple observations on the same pulsar, we only use the observation in which the pulsar's integrated intensity was brightest. It is pertinent to note, however, that in duplicate observations, the modulation statistics were reproducible. In only a few cases the energy distribution had a different classification in the fainter observation, typically transforming an \emph{\textbf{L}}-class pulsar to an \emph{\textbf{O}} or \emph{\textbf{U}}, and likely caused by the stronger influence of noise on the lower S/N observation.

\begin{figure}
\centering
\includegraphics[angle=90,width=0.47\textwidth,trim=19mm 14.2cm 8cm 4cm, clip]{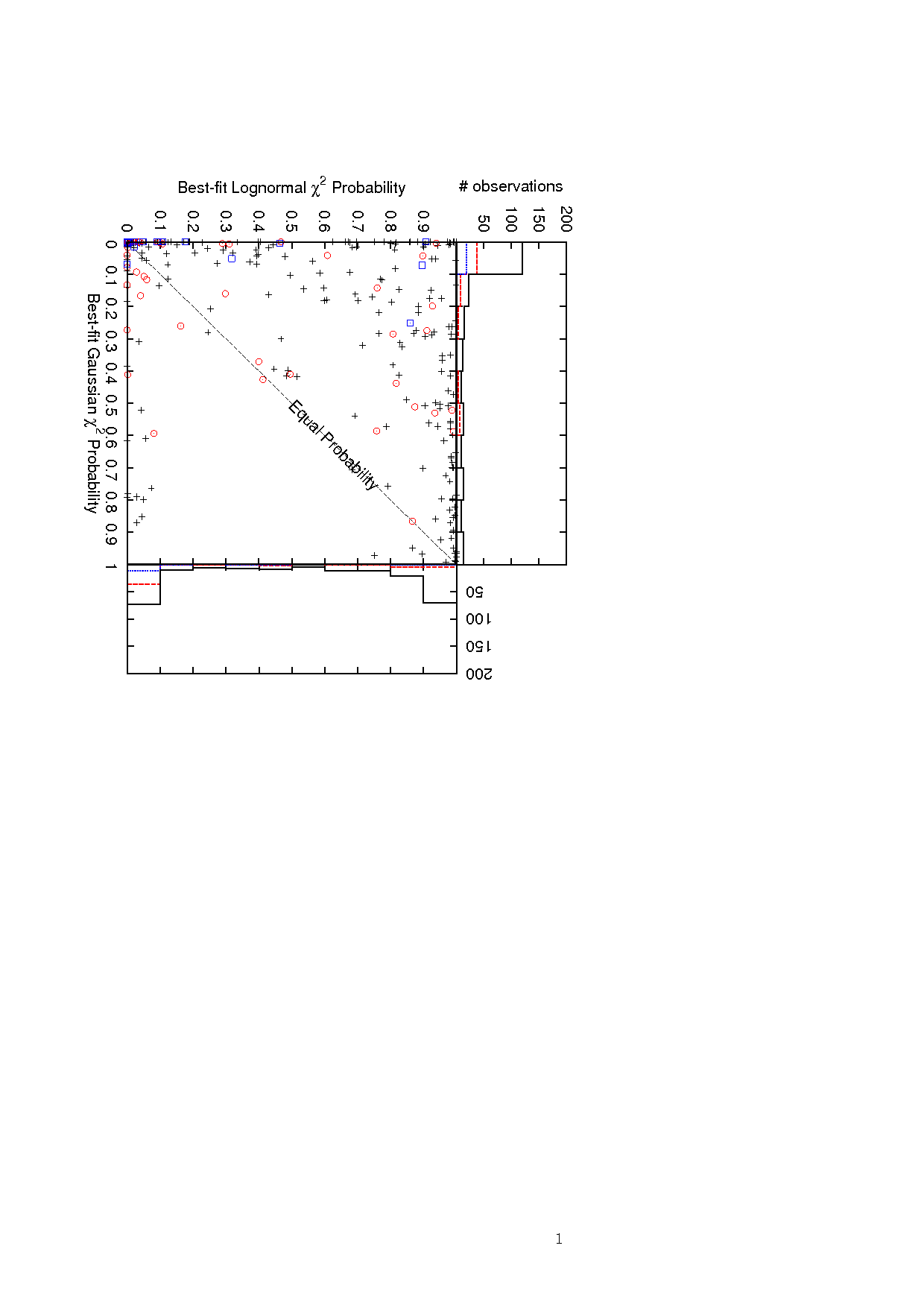}
\caption{A comparison of our best-fit distribution probabilities. Log-normal distributions are clearly favored. All objects classified as $G, L, U$ or $O$ (\S\ref{sec:eclass}) are shown as black crosses (solid line), while unclassifiable nulling pulsars are shown as red circles (dashed line) and multi-peaked distributions are shown as blue squares (dotted line). The upper and right panels show the integrated distribution of Gaussian and log-normal probabilities, respectively.}\label{fig:chicomp}
\end{figure}

\subsection{The distribution statistics of pulse energy} \label{sec:estats}
% SECTION OUTLINE:
%\begin{itemize}
%	\item Things seem to lean towards log-normal/asymmetric. The smooth tail of objects between $P=0$ and $P=1$ for both log-normal and gaussian can be due to a number of effects, discussed below in section \S\ref{sec:other}
%	\item Many fit log-normal, based on our threshold. We think the stats are not erroneous because the high $\se$ pulsars are close to either zero or $>$0.75 probability. What are the properties of these pulsars? With and without the ``unimodal'' ones.
%	\item There are some Gaussian ones. Their distributions seem valid. Do they have any particular physical parameters in common?
%	\item What does it mean that most are log-normal and some are gaussian?
%	\item ``Other''
%\end{itemize}

%While the distribution-fitting analysis resulted in 48\% of pulsar distributions that did not exceed our goodness-of-fit threshold for either a log-normal or Gaussian distribution (discussed further below),
More than one third of our classifiable sample was found to be above our threshold for agreement with a log-normal distribution. This is accented by Figure \ref{fig:chicomp}, in which we show a comparison of the best-fit probability for all pulsars. The general tendency of the energies away from a symmetric, Gaussian distribution here is pronounced. The origin of the tail of objects across all probabilities for both the log-normal and Gaussian trials is thought to be a low S/N effect, and is discussed below.
A primary target of this analysis was to determine whether pulsar energy is well-fit to a Gaussian or log-normal distribution, and if so, what distribution parameters are typical. We will focus momentarily on determining the parameters of the log-normal pulsars in our sample. We include unimodal objects in this discussion on the basis of their agreement with a log-normal distribution. The distribution of both $\sln$ and $\mln$ are qualitatively similar for the log-normal and unimodal sources (and a K-S test between the distributions does not support the null hypothesis).

\begin{figure}
\centering
\includegraphics[angle=270,width=0.47\textwidth,trim=3mm 10mm 8mm 1mm, clip]{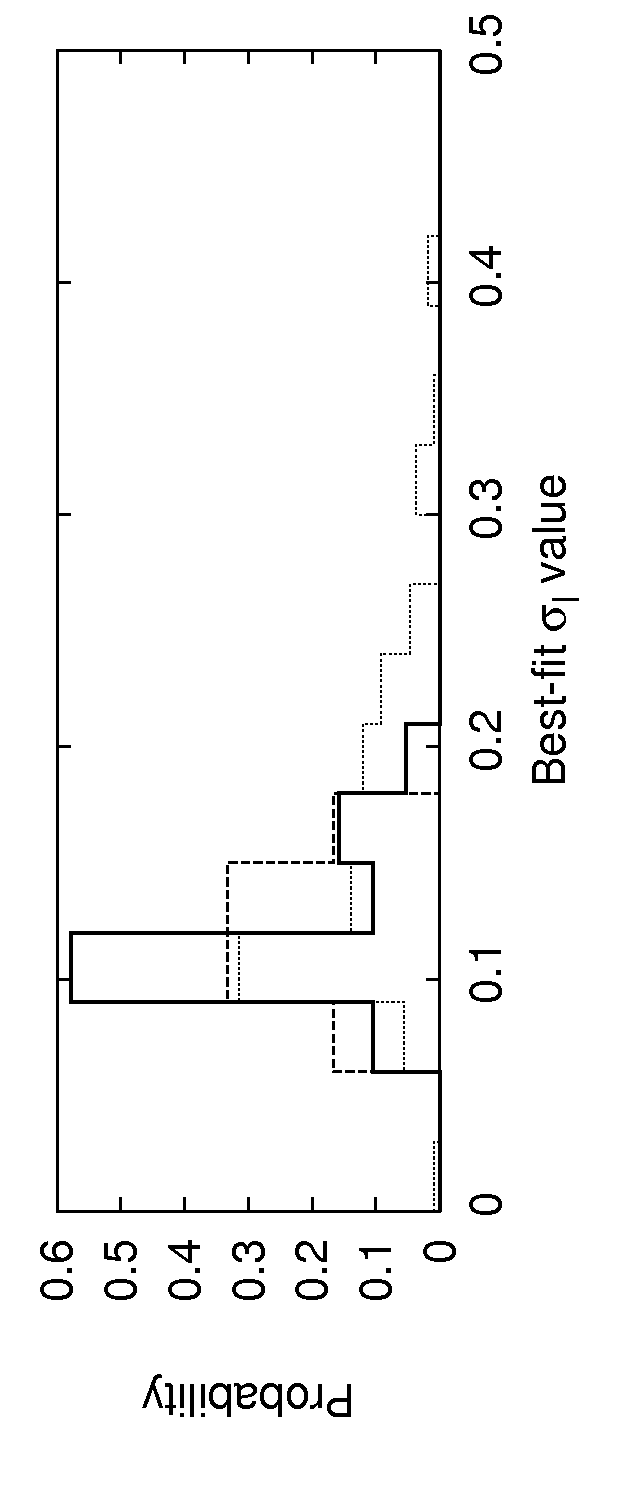}
\caption{The normalised best-fit $\sln$ distribution for strong-signal (thick dark line), non-nulling (dashed line), and all sources under the log-normal classification.}\label{fig:lnsigdist}
\end{figure}

\begin{figure}
\centering
\includegraphics[angle=270,width=0.47\textwidth,trim=3mm 10mm 8mm 1mm, clip]{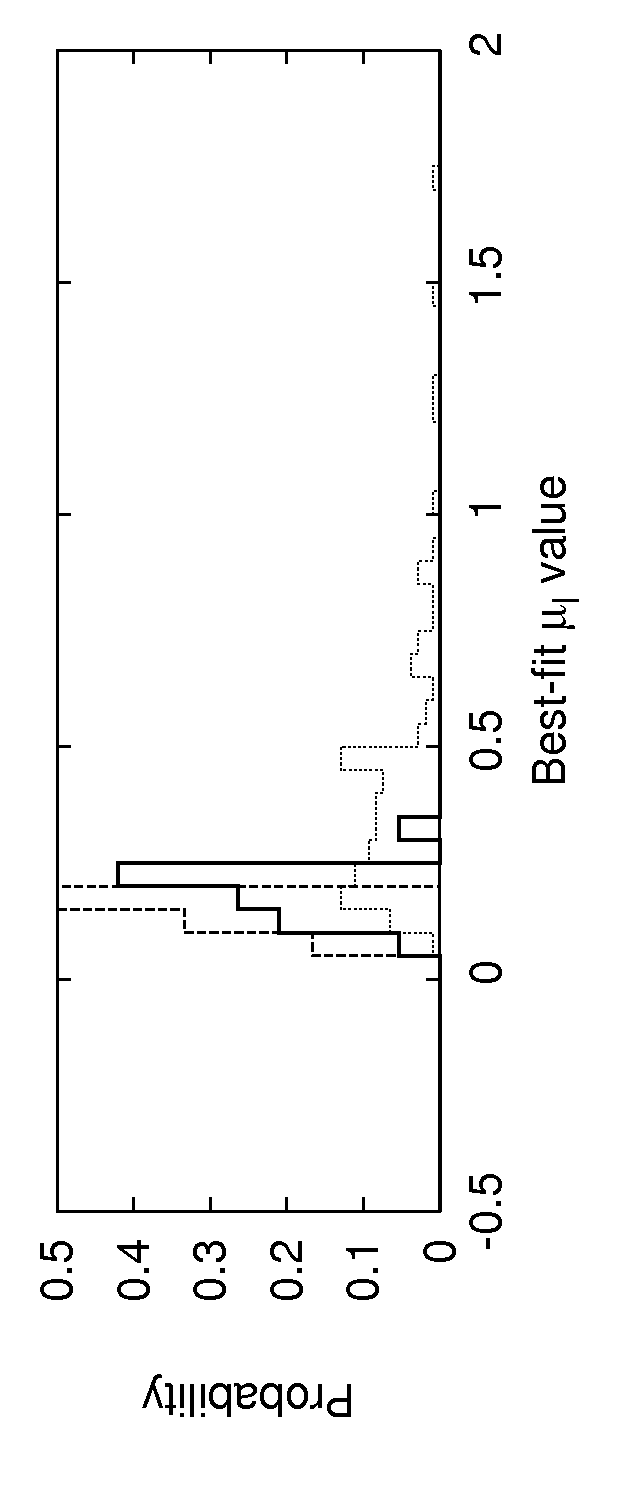}
\caption{As in figure \ref{fig:lnsigdist}, but reporting the best-fit $\mln$.}\label{fig:lnmudist}
\end{figure}

Care must be taken when considering the $\sln$ and $\mln$ results for our log-normal targets. Low S/N single pulse observations can lead to average single pulse energies which lie below the receiver noise (thus, we see e.\,g. only noise and the log-normal tail of the brightest pulses), and may limit our ability to identify null pulses. The presence of null and multi-peaked pulsars at high $\pln$ in Figure \ref{fig:chicomp} already indicate that multi-modal pulsars may contaminate the log-normal sample.  
%The multi-peaked population appears to be small and we do not expect a significant contamination from such pulsars. However, 
Unidentified nulling sources and low-signal measurements may potentially skew the log-normal parameter estimation, and we can see evidence of such an effect in an anti-correlation of low-$\se$ pulsars with $\mln$ in our data. To avoid contamination of our potential correlations by low S/N data, we measured the $\sln$ and $\mln$ distributions only for pulsars whose single pulses were on average detected with significance $\se>4$.
Figures \ref{fig:lnsigdist} and \ref{fig:lnmudist} compare the probability distribution of the best-fit $\sln$ and $\mln$, respectively, of the $\se>4$ objects to the distributions of non-nulling and all objects. The non-nulling sources, although numbering only 6, are completely unaffected by pulse nulling and provide a consistency check for the $\se>4$ distributions; a Kolmogorov-Smirnov test between the non-nulling and $\se>4$ pulsars do not support the null hypothesis for $\sln$ or $\mln$. The distribution of all sources is found to differ significantly from the $\se>4$ sources (supporting the null hypothesis at probabilities of 0.006 and $<$0.001 for $\sln$ and $\mln$, respectively). This is not thought to be a physical effect, but as previously stated is likely to be caused by errors in parameter estimation in the low-signal sample due to the influence of noise or unidentified nulling. For comparison, we report the mean and standard deviation of $\sln$ and $\mln$ for the three populations in Table \ref{table:slnmln}.

\begin{table}
\centering
  \caption{Average best-fit log-normal parameters for three subsamples of the pulsars classified as log-normal or unimodal. Note that the $\se>4$ sources provide the fiducial sample values. Variance of the sample's values is given in parentheses.}\label{table:slnmln}
%{\bf PUT $\mln, \mu_{G}$ HERE OR NO? NOT GENERALLY USEFUL BUT MAYBE...}
\begin{tabular}{rccc}
\hline
Sample & \textbf{$N$} & $\langle\sln\rangle$ & $\langle\mln\rangle$\\
\hline
\textbf{$\se>4$} & 19 & 0.11 (0.03) & 1.18 (0.07) \\
\textbf{Non-nulling} & 6 & 0.12 (0.03) & 1.13 (0.04) \\
\textbf{All} & 105 & 0.15 (0.07) & 1.50 (0.42) \\
\hline
\end{tabular}
\end{table}

Only three percent of our classifiable population were in agreement with a Gaussian distribution, of which only four objects (PSRs J0738--4042, J1507--4352, J1651--5222, and J1807--0847) had average single-pulse S/N of greater than four. Of all the observed and derived physical properties tested ($\tau_{\rm c}$, $B$, $P$, $\dot{P}$, dispersion measure, pulse width, and duty cycle), none stood out for these pulsars from the pulsars in the general population. Furthermore, they seem to share no characteristics in pulse shape or modulation, except that three of the four objects exhibit peaks in the $\rj$ modulation parameter in the centre of the profile. However, this is not a characteristic that is unique to these objects.
% as discussed in Section \ref{sec:modstats} below.
Both PSRs J0738--4042 and J1651--5222 exhibit intricate features in $\rj$, the former showing intensely-modulated emission on the trailing pulse edge, and the latter appearing to exhibit two emission modes of similar energy, and possible sub-pulse drift. It is possible that these two objects have been misidentified as Gaussian, but in fact contain several profile modes whose mean energy properties share similar values.

\begin{figure}
\centering

\subfigure[PSR\,J1048--5832]{\label{fig:1048}
\includegraphics[trim=0mm 0mm 0mm 0mm, clip,width=0.24\textwidth]{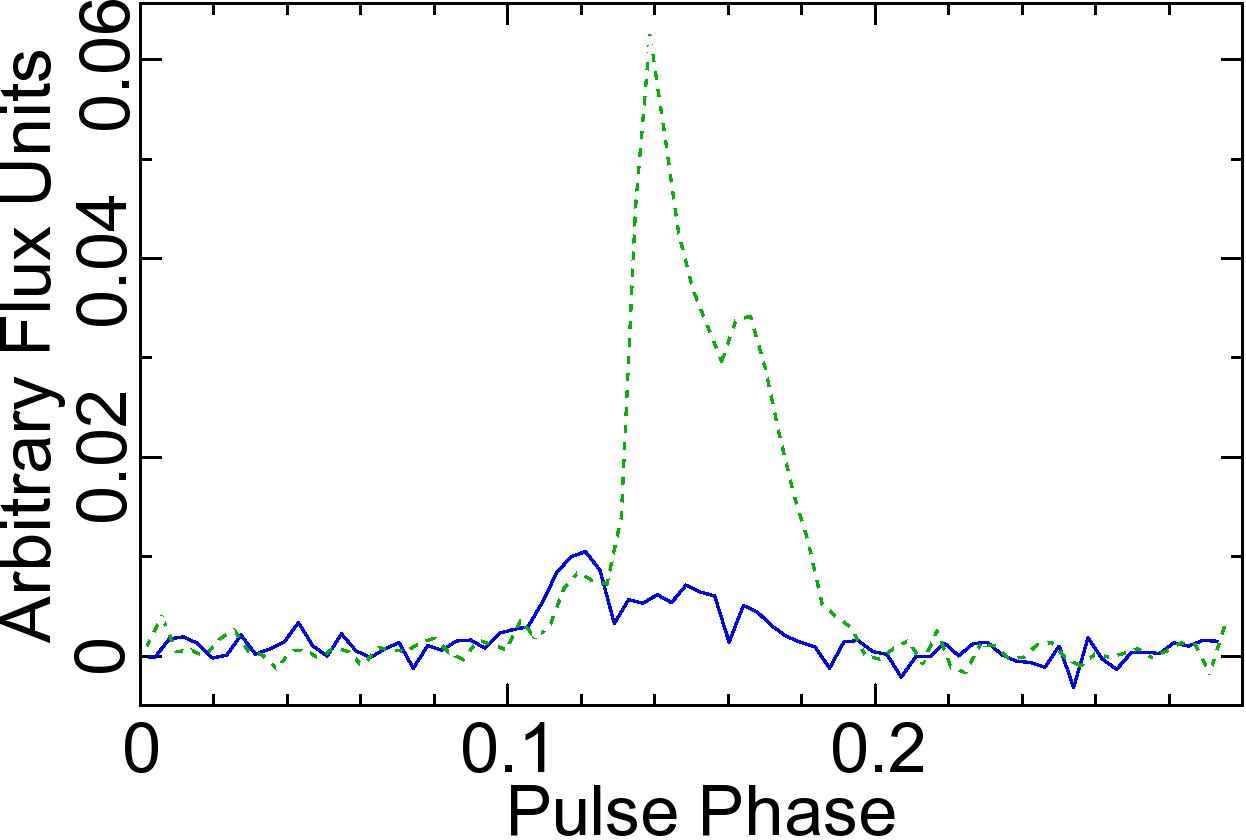} 
\includegraphics[trim=5mm 1mm 0mm 6mm, clip,width=0.24\textwidth]{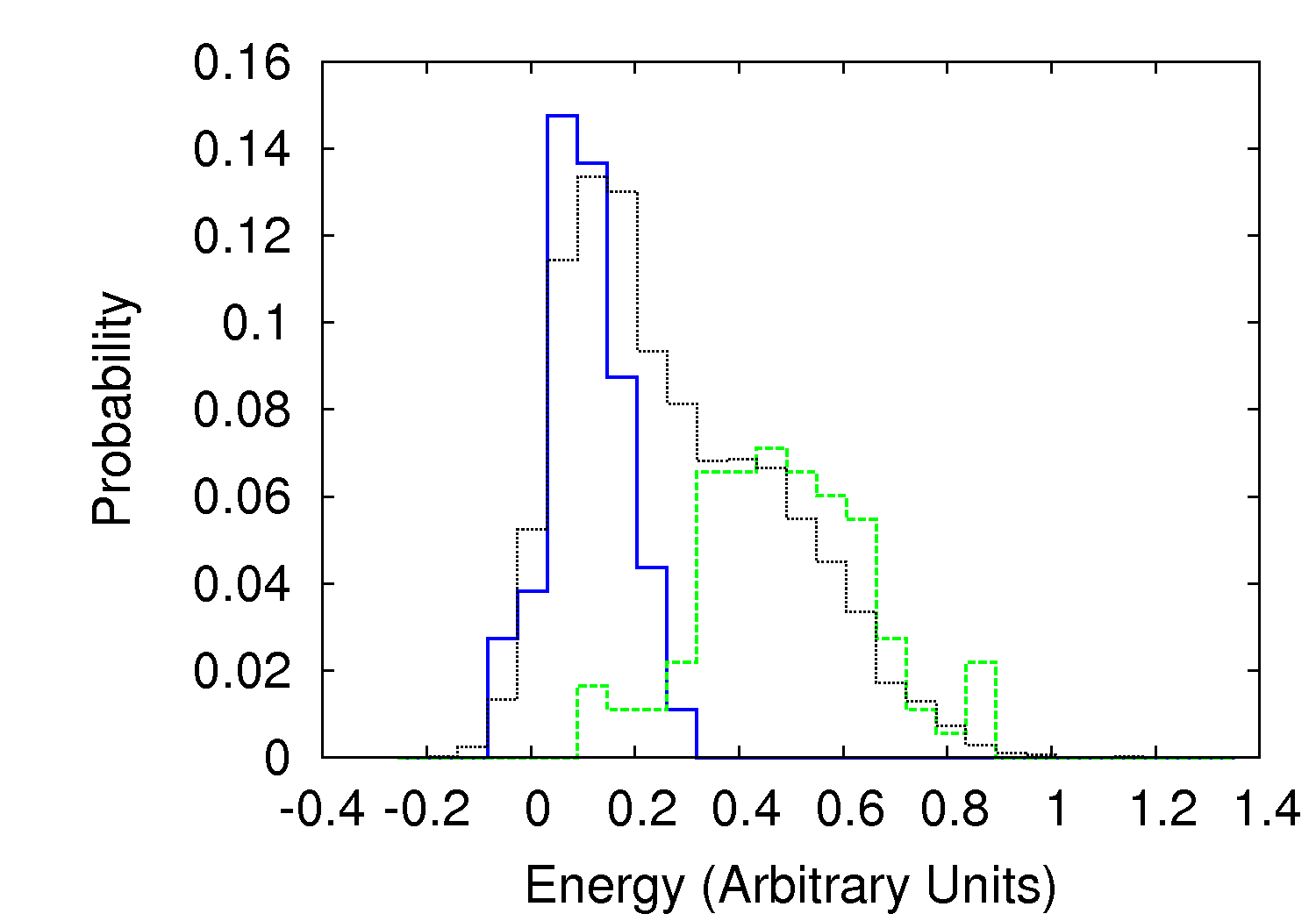}
}\quad
\subfigure[PSR\,J1900--2600]{\label{fig:1900}
\includegraphics[trim=0mm 0mm 0mm 0mm, clip,width=0.24\textwidth]{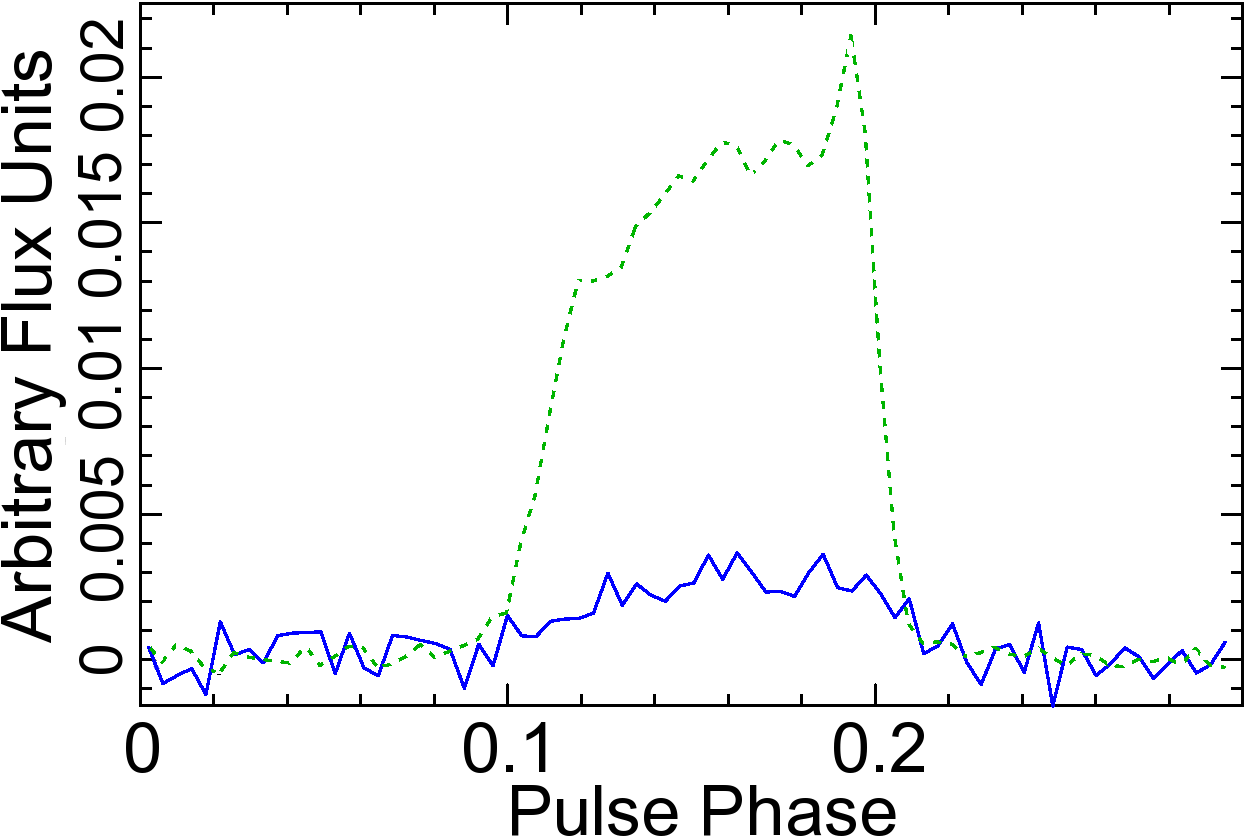}
\includegraphics[trim=5mm 1mm 0mm 6mm, clip,width=0.24\textwidth]{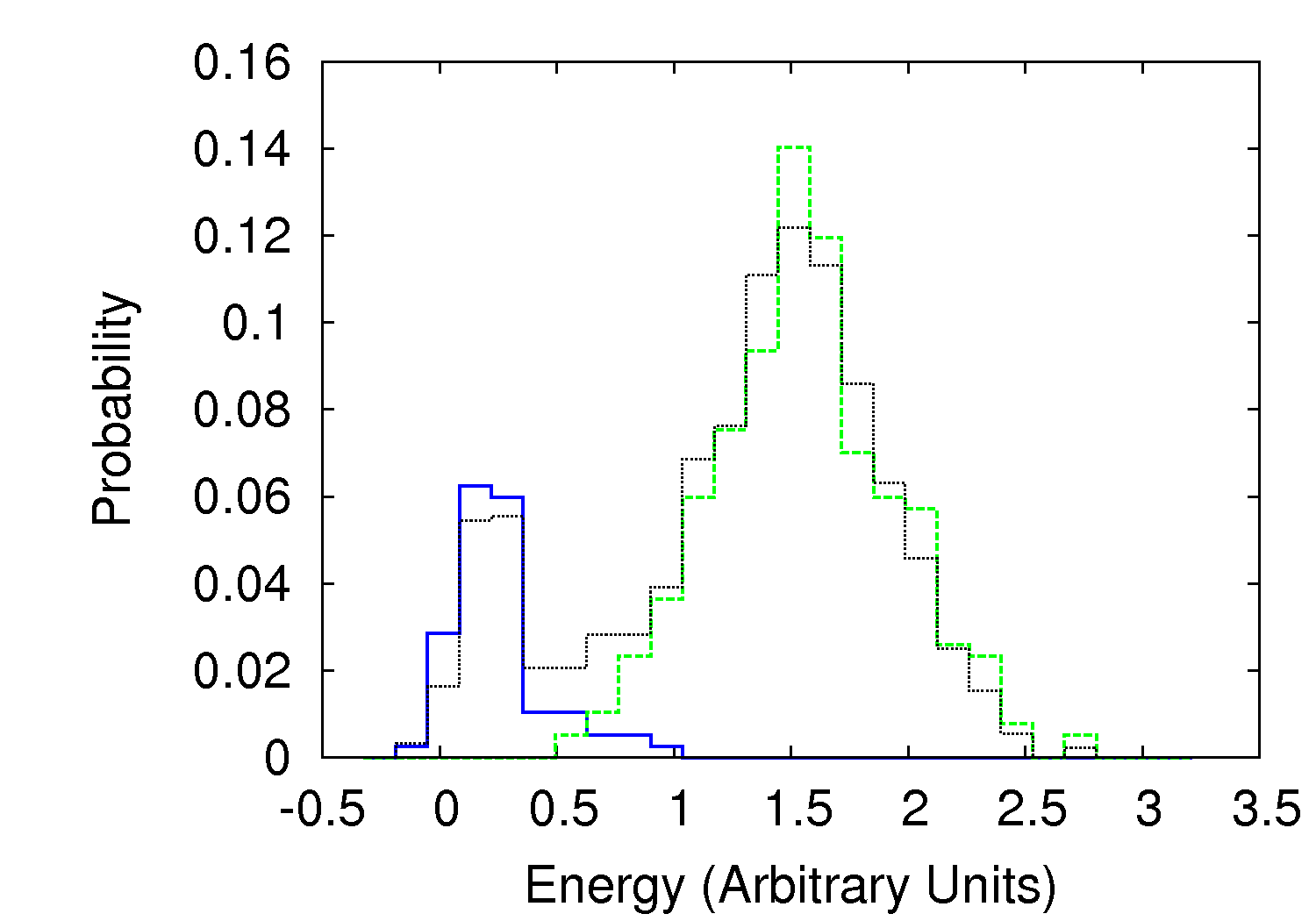} 
}
\caption{A view of the profile modes and their related energy distribution for two of our objects whose energy distributions were categorised as ``multi-peaked''. In all panels, the dashed green, solid blue, and dotted black lines correspond to the brighter mode, fainter mode, and all combined pulses, respectively. The left panels show the pulse profile integrated over a subset of pulses in each mode. The right panels show a mode-divided energy analysis as well as the integrated analysis. In both cases, two modes account fully for the multiple peaks identified in the energy distribution, and the non-zero mean of the off-peak distribution is clear.}\label{fig:moding}
\end{figure}

\subsection{Multi-modal energy distributions}\label{sec:bimodal}
% SUPPORTING THIS: wang et all (2007), esamdin (2005) although the latter of these is just 0826-34...
% What I need to discuss is that we see energy level changes although no nulls are apparent in all cases by eye. This might be a good way to a) identify energy reconfigurations and b) study the statistics of apparent nulls vs. mode changes, to know how many nulls are actually mode changes! 
% See nina's paper for a better worded description of what's happening in terms of current reconfigurations

A total of 18 pulsars in our sample had energy distributions classified as multi-modal. It appears that the majority of these multi-modal distributions are caused by mode changes; those with large relative energy peak differences exhibit mode changes that are visibly distinguishable in pulse stacks. We show two cases of this in Figure \ref{fig:moding}, in which each pulsar exhibits two profile configurations that correspond to a change in observed energy.\footnote{By visual inspection of the energy distributions of the two modes, it appears there might also be a change in energy distribution statistics accompanying the mode change. This would have fascinating implications, however we defer discussion on this until a more rigorous analysis can be performed.} It is likely that all multi-peaked objects in our sample are the result of such profile reconfigurations, even if they are not always readily identifiable in our pulse stacks (e.\,g. due to faint emission and barely-resolved profiles). For some pulsars, we cannot rule out that a transient component (e.\,g. giant bursts) on an otherwise steady profile is causing the second peak. We find it worthy of explicit mention that the inspection of energy distributions appears to be a straight-forward way to identify mode changing in many pulsars, in which it might not be obvious from an inspection of only a pulse stack.

A clear ramification of the energy difference associated with mode changes is that some nulling pulsars may be exhibiting mode changes in which either the energy state drops below an observation's noise level (distinct from cessation of emission), or the beam configuration changes sufficiently such that no sub-beams are aimed at Earth. This has been previously suggested \citep[e.\,g.][]{wangetal07,nullsaremodes}, and is supported by the faint emission seen in some pulsars after the integration of many ``null'' pulses. For instance, PSR\,J1900--2600 (Fig.\,\ref{fig:1900}) was previously identified as a nulling pulsar with a 10-20\% nulling fraction \citep{ritchings76,1900nulls}, which is approximately the fraction of pulses we observe in the low-energy mode. Our data's contributions to the ``nulls are mode changes'' hypothesis are three: 1) mode changing appears to be fairly prolific (6\% of our whole sample had \emph{discernible} multiple non-null energy peaks), 2) we observe a range of changes in mean integrated energy value, implying that some such pulsars could be misidentified as nulling or unimodal, thus the mode-changing population is probably larger, and 3) substantial reconfigurations may be more common than minor ones, given the 69 nulling pulsars and 18 multi-modal objects in our sample. Three of our nulling pulsars exhibit multiple non-null peaks, thus may have multiple mode changes.

%J1048 is ``off'' for almost exactly 10 periods each time!

%One problem maybe worth a mention here: Bi/multimodality creates asymmetry in the energy distribution, which if not clearly resolved could ``fool'' a fit of energy distribution into leaning towards log-normality. Point out the stats on that (which should be left in, in table \ref{table:stats}) to accent that fact.

\section{Modulation Statistics of Pulsars}\label{sec:modstats}
Here, we discuss several distinct topics relating to pulse-to-pulse modulation in pulsars: Section \ref{sec:moddist} presents the modulation values across our full sample, characterising the range of pulse-to-pulse modulation statistics of the general pulsar population. Section \ref{sec:pdist} discusses the phase-dependent location of modulation relative to the total intensity shape of the pulsar's profile in an attempt to understand if and how bursty (i.\,e. high-$\rj$) emission relates to the underlying pulsar beam shape. Finally, Section \ref{sec:correlations} describes correlation tests between the modulation parameters $R$ and $m$ with physical pulsar parameters.

In Table \ref{table:stats}, we report three indicators of pulsar modulation: the maximum on-pulse $\rj$ value, the minimum on-pulse $\mj$ value, and the S/N of the brightest single pulse detected in the blind single pulse search, when these measurements are significant for a pulsar. We follow the significance threshold for $\mj$ used by \citealt{patricketal06} and \citealt{jenetgil2003}, in which the S/N of the integrated pulsar profile must be greater than 100.
Because off-pulse values of $\rj$ reflect the data's radiometer noise properties, as previously noted, this statistic is only considered significant when the on-pulse peak $\rj$ value is more than 4 times the standard deviation of $\rj$ values in the off-pulse profile. Note that the maximum single pulse search S/N should not necessarily correlate with $\mj$ or $\rj$ because they are calculated at a fixed time-sampling, whereas the single pulse search utilized a box-car search to fit for ideal pulse width.

\subsection{Distribution of modulation parameters}\label{sec:moddist}
\begin{figure}
\centering
\includegraphics[angle=270,width=0.48\textwidth,trim=5mm 10mm 0cm 0mm, clip]{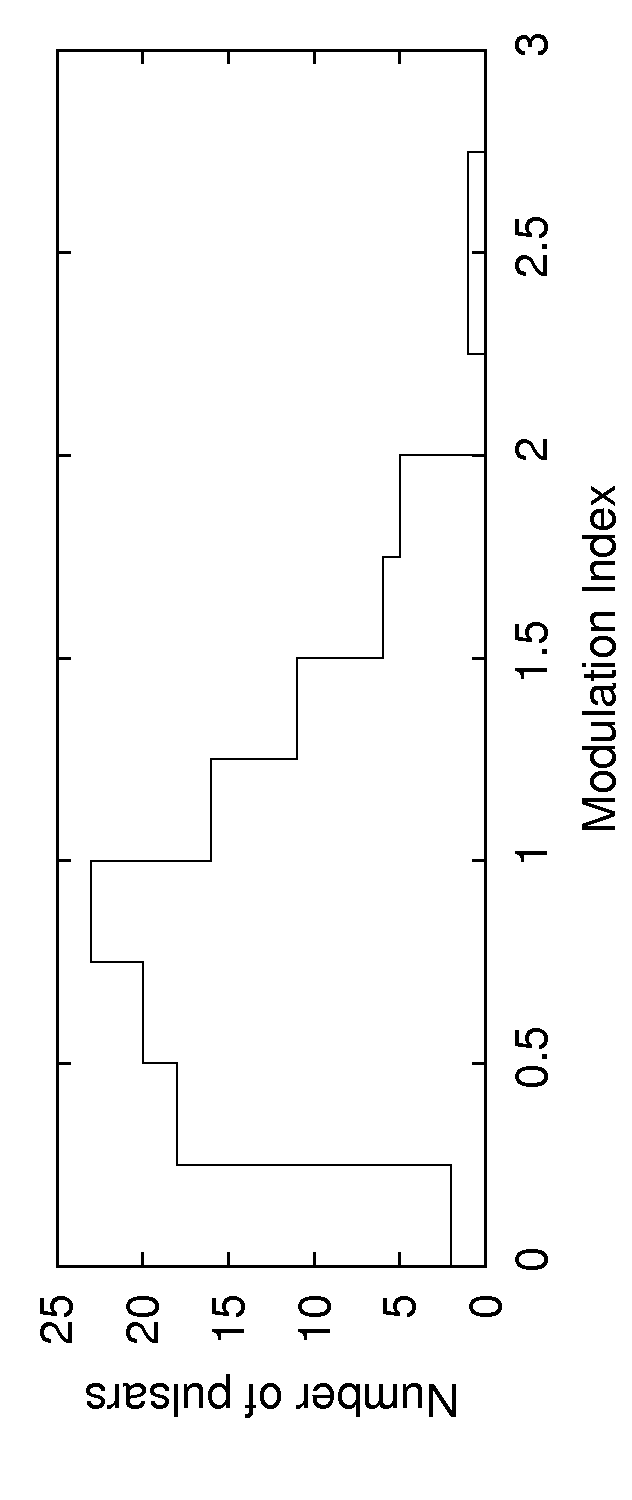}
\caption{The distribution of minimum $\mj$ value for the 103 pulsars with S/N$_{\rm int} \geq 100$, as discussed in \S\ref{sec:moddist}}\label{fig:mdist}
\end{figure}

The distribution of pulsars' minimum modulation index, $m$, provides a direct empirical snapshot of the pulsar population's typical modulation properties. Figure \ref{fig:mdist} shows the distribution of $m$ for our sample.
While we do not distinguish various drift phenomena as in \citet{patricketal06}, we can compare our results to theirs. In the Weltevrede et al. study, $m$ was measured using a longitude-resolved power spectral technique rather than direct computation. While our distribution agrees in peak value, ours is moderately broader, and more heavily weighted towards higher $m$ values than that of Weltevrede et al. This slight difference is possibly attributed to the difference in technique for mitigation of scintillation's contribution to $m$. While the Weltevrede et al. technique removed any low-frequency modulation (thus in addition to the $m_{\rm ISM}$ contribution,  potentially removing some modulation attributable to the pulsar itself), our mitigation may have included an erroneous estimate of $m_{\rm ISM}$ due to errors in the \citet{ne2001} electron density distribution model. We would expect the former point to most strongly contribute to the observed effect.

\begin{figure}
\centering
\includegraphics[angle=270,width=0.48\textwidth,trim=5mm 10mm 0cm 0cm, clip]{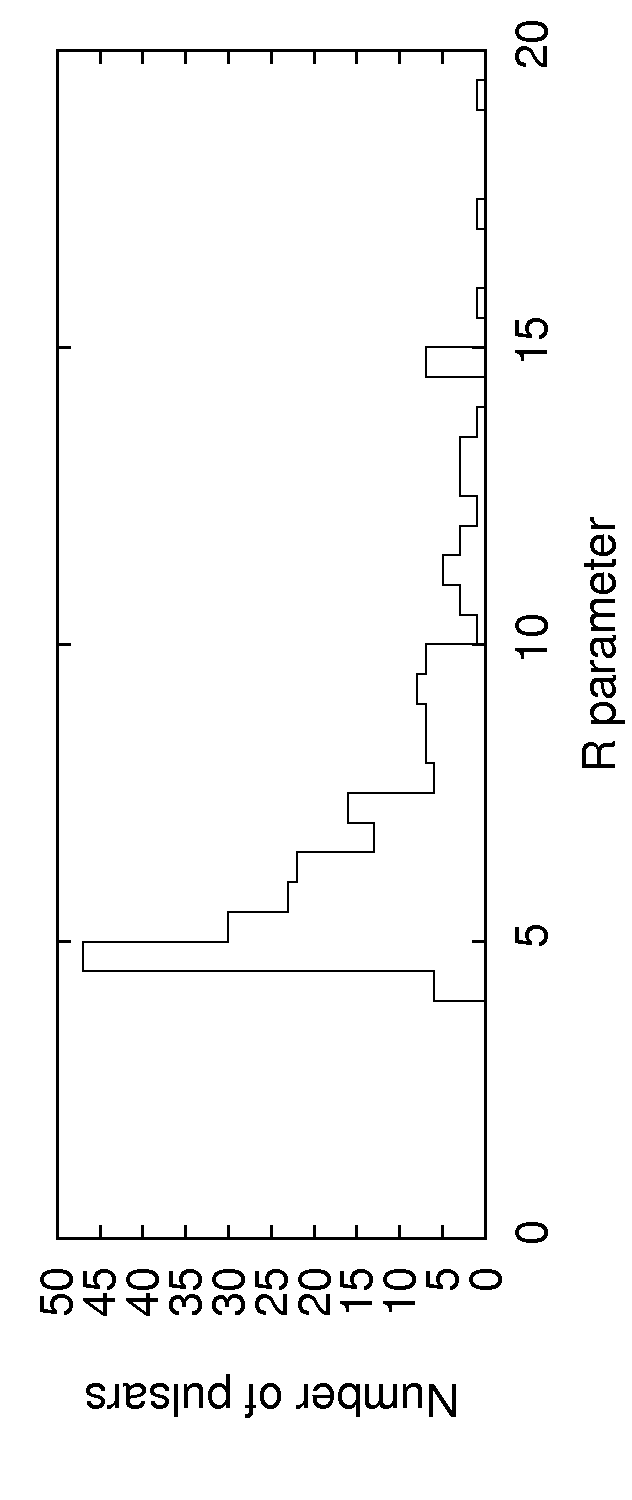}
\caption{The distribution of maximum $\rj$ value for the 222 pulsars for which this parameter was significant.}\label{fig:rdist}
\end{figure}

In Figure \ref{fig:rdist} we provide the $R$-parameter distribution. As previously noted and discussed further in \S\ref{sec:correlations}, the $R$-parameter distribution cannot be taken at face value to be an ``intrinsic'' modulation distribution due to its strong dependence on Gaussian statistics and mean single-pulse flux.
However, note that a measured $\rj$ value represents signal inconsistent with Gaussian variance; thus, these pulsars exhibit phase-resolved, sporadically-varying emission. If (both integrated and phase-resolved) pulsar energy distributions are indeed log-normal, this result is not entirely unexpected.  
We note that in observations of increasing sensitivity, $\rj$ values particularly of pulsars where the single-pulse mean is hiding in the noise (e.\,g. deep nulling pulsars and RRATs) will increase.
%(note, only 7 of 74 pulsars with $\se>4$ did not have a significant $R$ measurement (but more at lower SNR?)
Additionally, the observed maximum $\rj$ will scale with a sample's observing length, consistent with the probability distribution of emission energy. We thus expect that if the $\sln$ and $\mln$ values presented in Section \ref{sec:estats} hold for the full population, the distribution shown in Figure \ref{fig:rdist} would shift to higher values and perhaps broaden slightly, were our observing length and/or sensitivity increased.

%\begin{figure}
%\centering
%\includegraphics[angle=270,width=0.48\textwidth,trim=5mm 10mm 0cm 0cm, clip]{figs/pulsewid.png}
%\caption{The distribution of offsets between peak $\rj$ value and peak intensity, as a function of pulse width fraction for all objects (solid line) and pulsars not identified as nulling (dotted line).}\label{fig:roffs}
%\end{figure}

\begin{figure*}
\centering
\subfigure[PSR\,J0738--4042]{
\includegraphics[angle=270,width=0.4\textwidth,trim=5mm 5mm 3mm 0cm, clip]{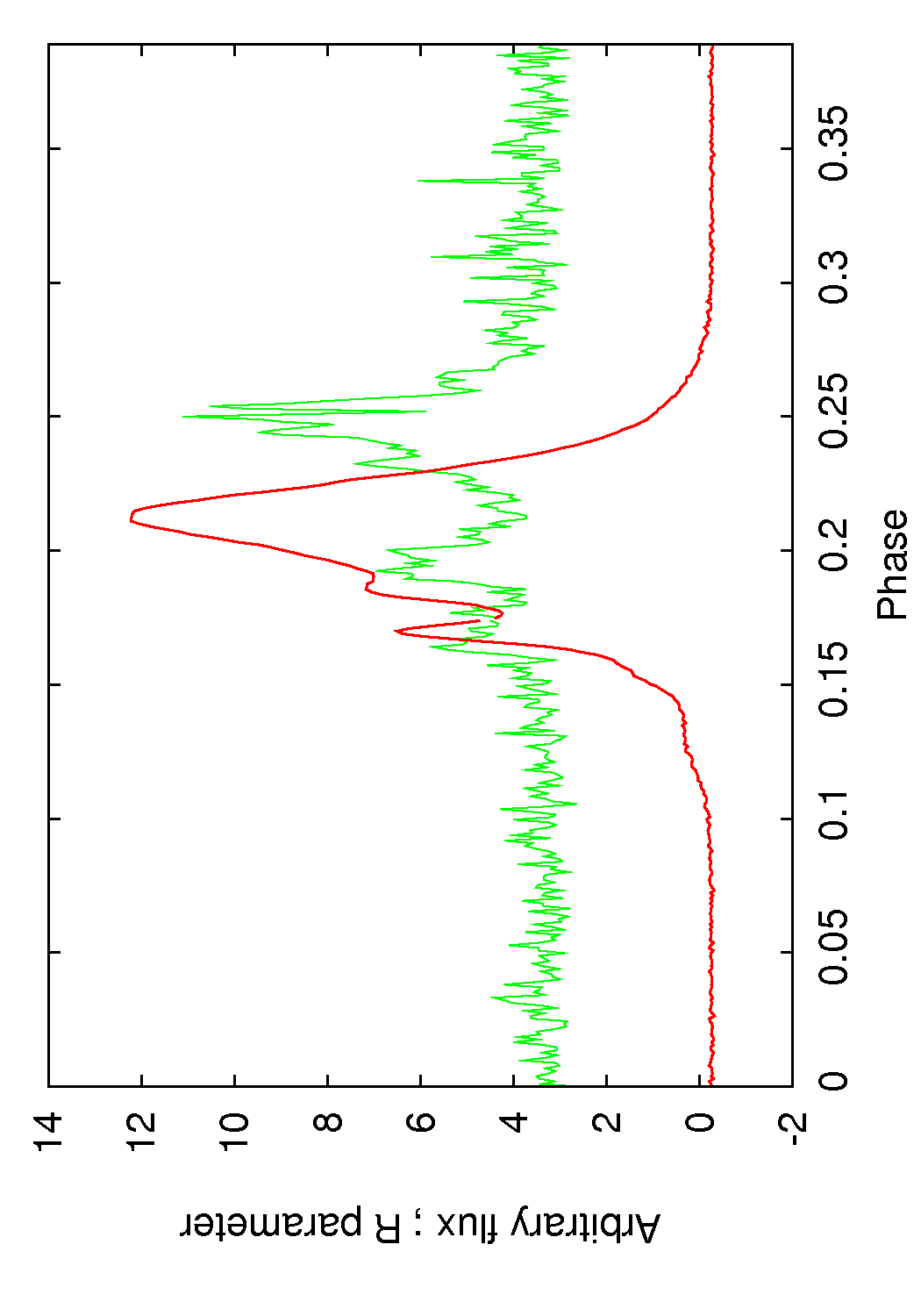}
}\quad
\subfigure[PSR\,J0934--5249]{
\includegraphics[angle=270,width=0.4\textwidth,trim=5mm 5mm 3mm 0cm, clip]{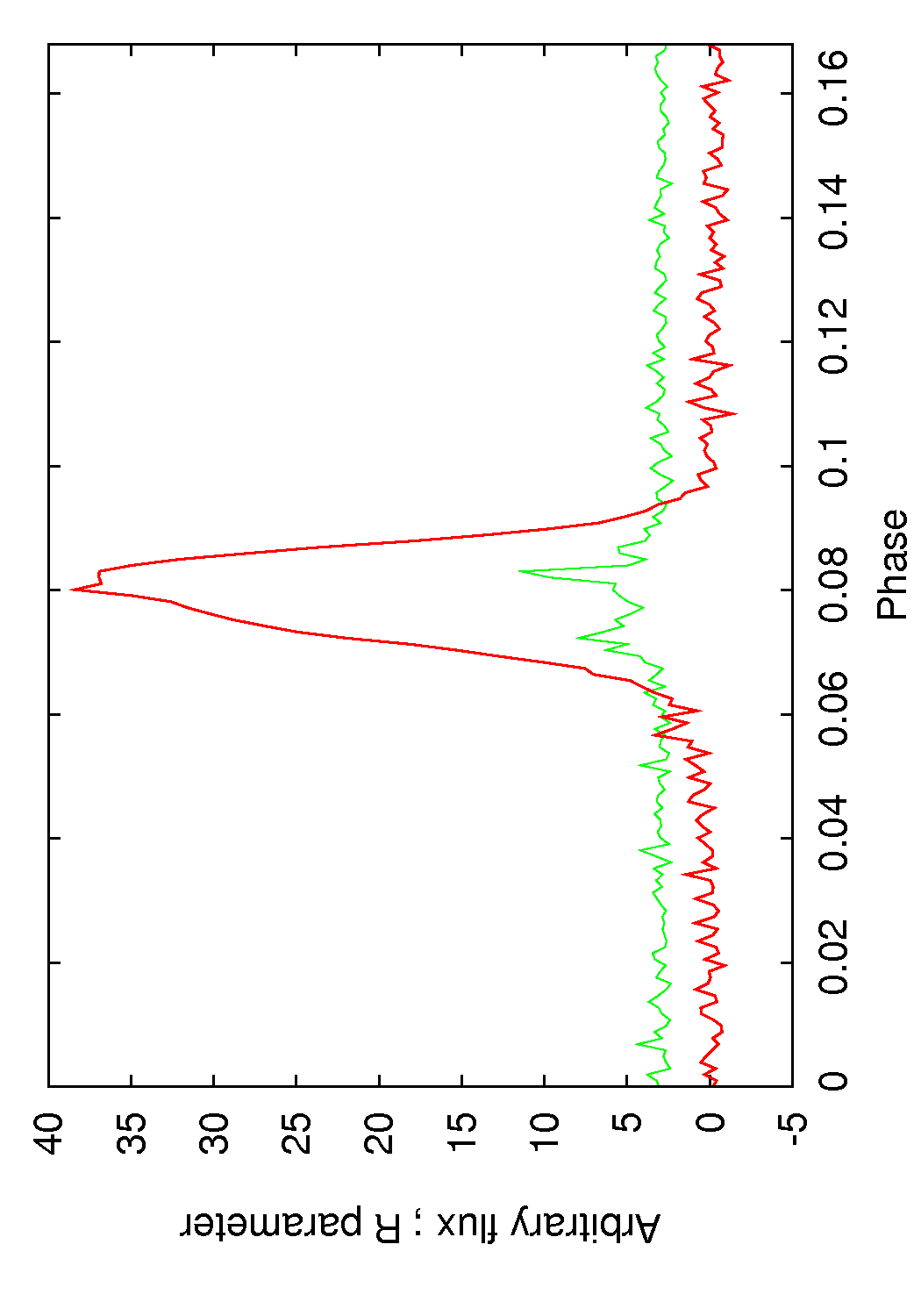}
}
\subfigure[PSR\,J1839--1238]{
\includegraphics[angle=270,width=0.4\textwidth,trim=5mm 5mm 3mm 0cm, clip]{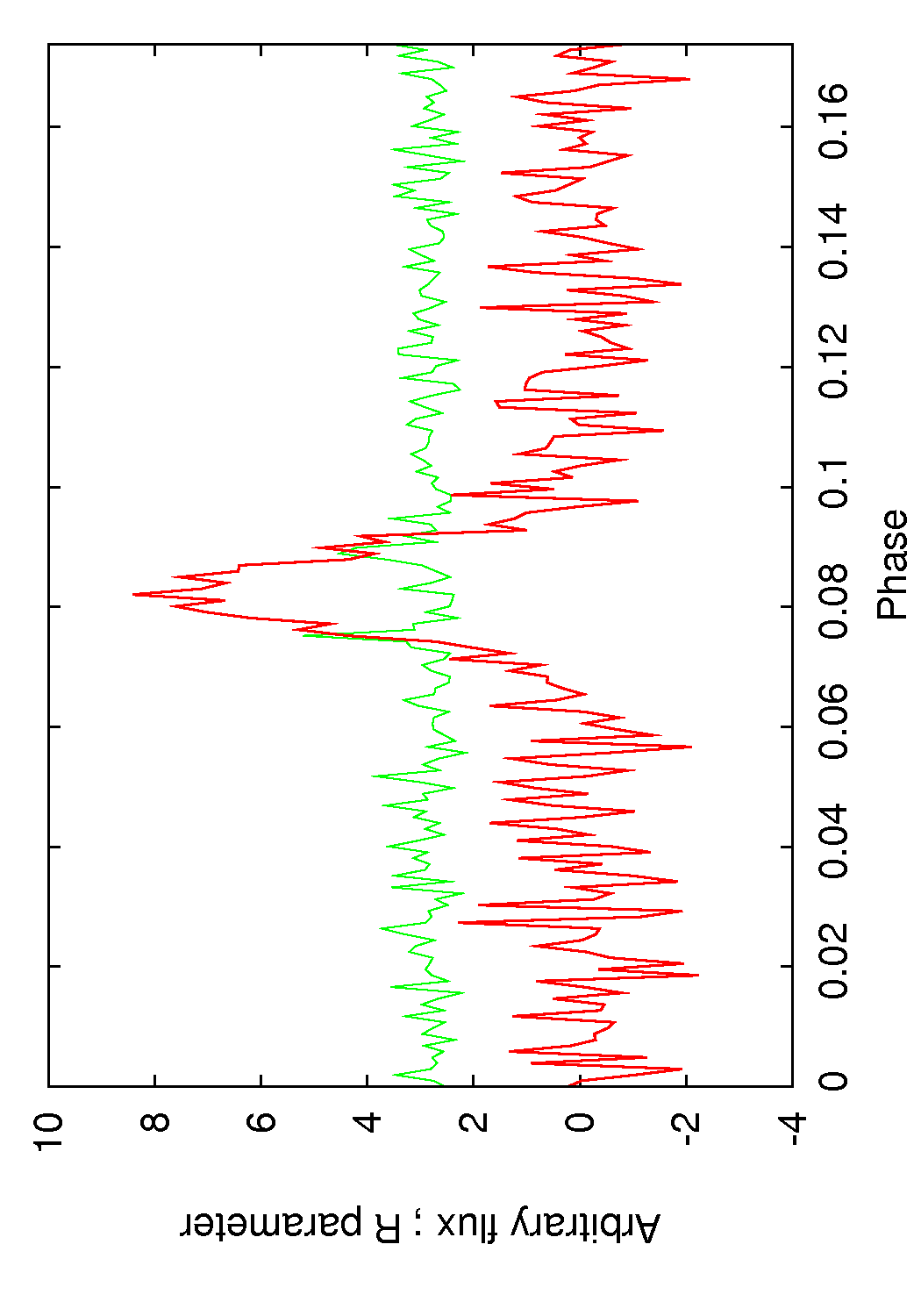}
}
\subfigure[PSR\,J1852--0635]{
\includegraphics[angle=270,width=0.4\textwidth,trim=5mm 5mm 3mm 0cm, clip]{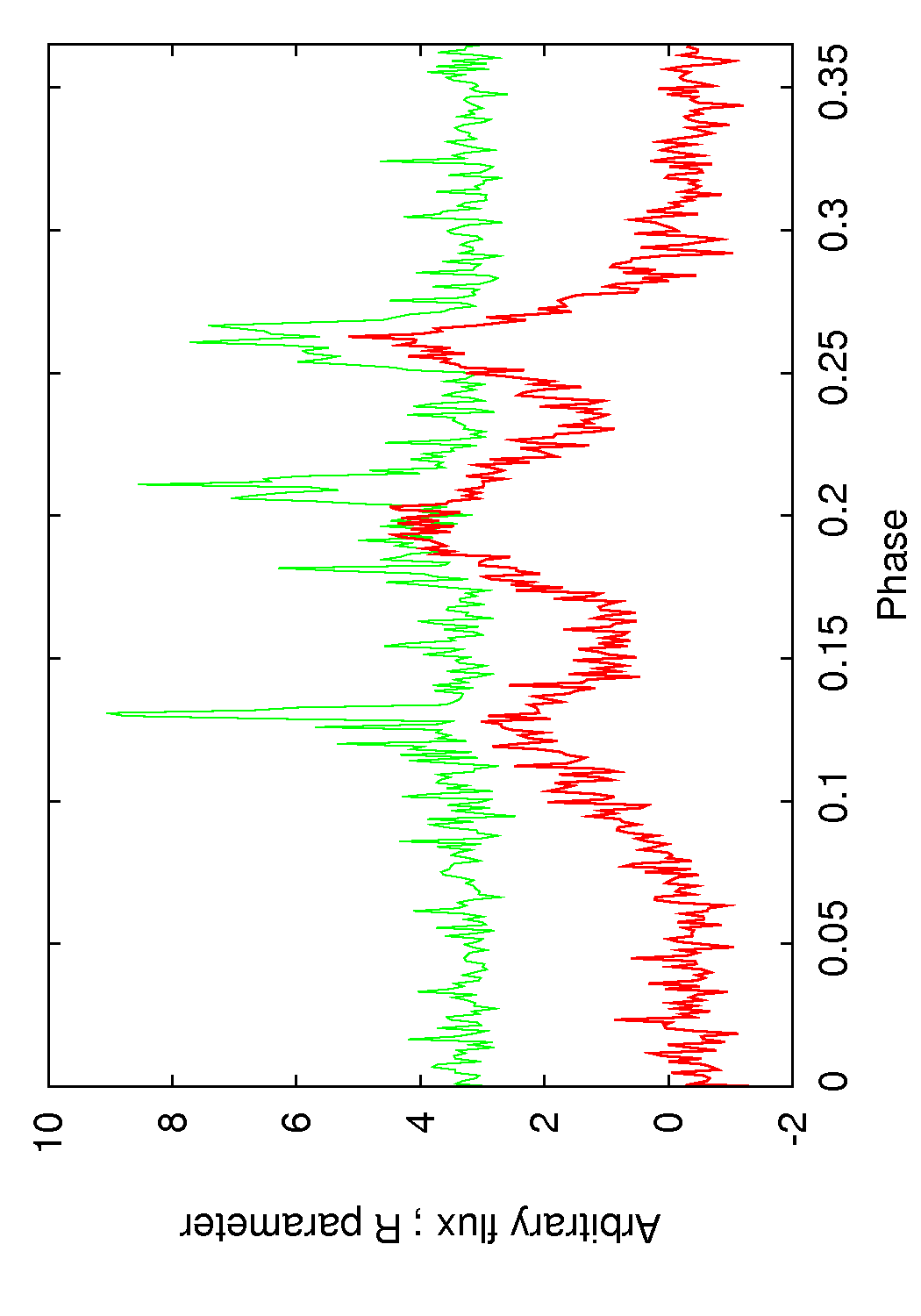}
\label{fig:1852}
}
\caption{An expanded view of various $\rj$ component profiles (thin green line) and their corresponding integrated intensity profile (thick red line). All of these profiles exhibit leading and/or trailing sporadic pulse components, primarily flanking local maxima in total intensity.}\label{fig:rexamples}
\end{figure*}

\subsection{Profile dependence of modulation}\label{sec:pdist}
It has been noted in the literature that ``core'' profiles (as defined by the profile classification scheme of \citealt{rankinI}) are both less modulated than ``conal'' profile components, and do not null. It has also been reported that giant pulse phenomena occur typically on the leading or trailing edge of pulsar profiles (certainly, the persistent modulation appears to be higher at pulse edges; $\mj$ rises at the leading and trailing pulse edges for nearly the entirety of our sample). The $R$-parameter enables sensitivity to phase-resolved, sporadic emission behaviours. To explore the typical location of such emission and its relationship, if any, to integrated intensity profiles, we inspected each pulsar's total intensity and $\rj$ profiles (sample $\rj$ profiles are shown in Fig.\,\ref{fig:rexamples}, and all $\rj$ profiles are shown in the online Figure; see material referenced in Appendix \ref{sec:appendix}).

Persistent multi-phase features in $\rj$ appear to come in two types: broad, diffuse features that in many cases follow the rise and fall of the integrated intensity, and narrow features which have no pronounced counterpart in the total intensity profile thus presumably correspond to very sparse outbursts. The phase-dependence of narrow $R$-parameter peaks varies vastly from pulsar to pulsar, however in many objects, narrow $\rj$ features appear on the edge of (leading or trailing) a \emph{local} maximum in integrated intensity (not necessarily the brightest beam component). In some, the $\rj$ profile is dual-peaked, with peaks falling on either side of the integrated profile. Examples of these are shown in Fig. \ref{fig:rexamples}. This is suggestive of a sporadic sub-beam-edge effect, however as we do not have sufficient information to break down our profiles into conal or core components, we cannot say whether this effect is distinct to one profile geometry. Note, however, that in some pulsars the modulation does peak at the same phase as the integrated profile (in fact, PSR\,J1852--0636 as shown in Figure \ref{fig:1852} exhibits contemporaneously-peaking modulation and intensity profiles for the outer sub-beams, but offset modulation peaks for the centre beam).

Finally, the variation of $\rj$ values across a profile indicates an important point that will be discussed further in Sections \ref{sec:giants} and \ref{sec:interps}, which is that the energy distribution (i.\,e.  log-normal/power-law/gaussian/etc. classification and distribution parameters) can be phase-dependent.

\subsection{Correlation of modulation parameters with other neutron star parameters}\label{sec:correlations}
We performed Kendall's tau correlation tests for the minimum $\mj$ and maximum $\rj$ against all basic pulsar physical parameters: age, magnetic field strength, energy loss rate, $P$, $\dot P$, and DM. We found no significant correlations that were not directly accountable by sample selection effects. $m$ was found to correlate (in some cases, anti-correlate) with several parameters, most strongly with characteristic age and $\dot{E}$. However, we attribute all of those correlations to the strong anti-correlation between $m$ and integrated S/N (Kendall's $\tau$ statistic: -0.55; probability $P_{\tau}<0.000001$), which can induce correlations with $m$ due to our fixed observation length (this ``correlation'', as with the $R$-parameter/DM correlation below, is thought to be primarily the result of the low-weighted distribution of $m$ and the few objects with strong integrated signal; the distribution of $m$ at different signal intervals does not differ).
%, and therefore higher signal for shorter-period pulsars that have many rotations in an observation.
The correlations were not significant when restricting the tests to pulsars with an integrated S/N between 100 and 400, indicating that these correlations were induced by the brightest $\sim$20 objects.

We measured no significant correlations between $\mj$ and any of the complexity parameters of \citep{jenetgil2003}, in agreement with \citet{patricketal06}. Correlations between $m$ and the complexity parameters are predicted to be stronger when considering $m$ strictly from core pulse profiles \citep{jenetgil2003}; it is possible that if any correlations exist within this data, they are diluted by our lack of information about profile type and beam viewing angle. Potential errors in the NE2001 electron density model, leading to an incorrect treatment of scintillation's contribution to $m$, could also contribute to weakening a correlation. Thus, with the available information, our sample is unable to rule out any of the proposed theories with these correlation tests.

One correlation found with maximum $\rj$ warrants brief discussion; the maximum $\rj$ was weakly anti-correlated with dispersion measure ($\tau=-0.27$; $P_{\tau}<0.000001$). We interpret this primarily as the naturally low-weighted distribution of maximum $\rj$ (seen in the low-DM pulsars) and the fewer number of pulsars at high DM. However, pulse smearing and scattering may also dampen $R$-parameter values at high DM.

\section{General Discussions}\label{sec:remaining}
Here we address three remaining points of discussion that arose from our analysis. Section \ref{sec:giants} discusses physical motivators for the definition of the ``giant pulse'' phenomenon in pulsars based on our energy distribution and pulse-to-pulse modulation measurements. We furthermore indicate how our analysis may indicate giant pulse activity occurring in several pulsars. Section \ref{sec:interps} discusses the implication of our results for the net pulsar energy circuit, paying particular attention to a discrepancy between the narrow range in integrated single-pulse energy values versus the large range in phase-resolved bursty emission. We also draw on the results of interpulse-pulsars in this discussion. Finally, in Section \ref{sec:weirdos}, we point out peculiar behaviours observed in several pulsars that were identified in the course of our analysis.

\subsection{Giant pulses vs.~log-normal pulses}\label{sec:giants}
The definition of ``giant pulse'' has varied in previous analyses, with some authors defining the term as any pulse with a flux more than ten times the average flux at that phase, and others differentiating giant pulses by their power-law energy distributions. In our analysis, the former definition translates directly to the specification $R>10$. This condition is not uncommon in this data set, and furthermore the R parameter's continuous distribution over a broad range indicates that this differentiation of ``giant pulse'' is entirely arbitrary. While it is certainly a convenient definition, if many pulsars are indeed log-normally distributed, no physical distinction (except for small variations in $\sln$) should exist between high and low-$\rj$ pulsars. We therefore support the latter definition of ``giant pulse,'' which in addition denotes a clear difference in underlying plasma processes.

As we have previously noted, a significant measurement of $\rj$ implies the presence of non-Gaussian statistics in phase bin $j$, and does not strictly differentiate between what non-Gaussian distribution is causing the heightened $\rj$. For the pulsars with significantly measured $\rj$ values, we have an insufficient number of pulses in our data to perform an assessment of whether the phase-resolved energy distributions are caused by a pure log-normal distribution, or by the log-normal plus power law tail that is exhibited at giant-pulsing phases in some pulsars. However, studies of these high-$R$-parameter objects over a longer timescale could provide the data necessary to differentiate pulsars with broad phase-resolved log-normal distributions from power-law-distributed giant pulses as the cause of the intense modulation \citep[see e.\,g.][]{karuppusamy11}.

Although the broad time resolution used in our observations would dampen the intensity and prominence of giant micropulses, we can check for an indication of such activity 
by inspecting the data for very narrow (i.\,e. unresolved in phase), significant features in $\rj$. Several pulsars shows clear potential signs of such an effect: PSRs J0726--2612, J1047--6709 (the small, narrow feature preceding the main pulse), %has a crazy braod energy dist!
J1759--1956, and J1801--2920 (see Appendix A). 
%While the energy distribution classification for PSR\,J1801--2920 resulted as log-normal, there is some evidence at high energy of an overabundance of pulses, possibly indicative of a power-law tail caused by the narrow $\rj$ feature preceeding the main pulse. Fig\,\ref{fig:rdist} shows an expanded view of the $R$ and integrated profile for this pulsar.

%Also, we integrated over the whole pulse profile. How would that effect our results?
%What do our results mean? Can we interpret them in the light of some particular theory?!
%How do the gaussian things fit in?
%power-law: non-linear wave collapse

\begin{figure}
\centering
\includegraphics[angle=270,width=0.47\textwidth,trim=0mm 0mm 1mm 0mm, clip]{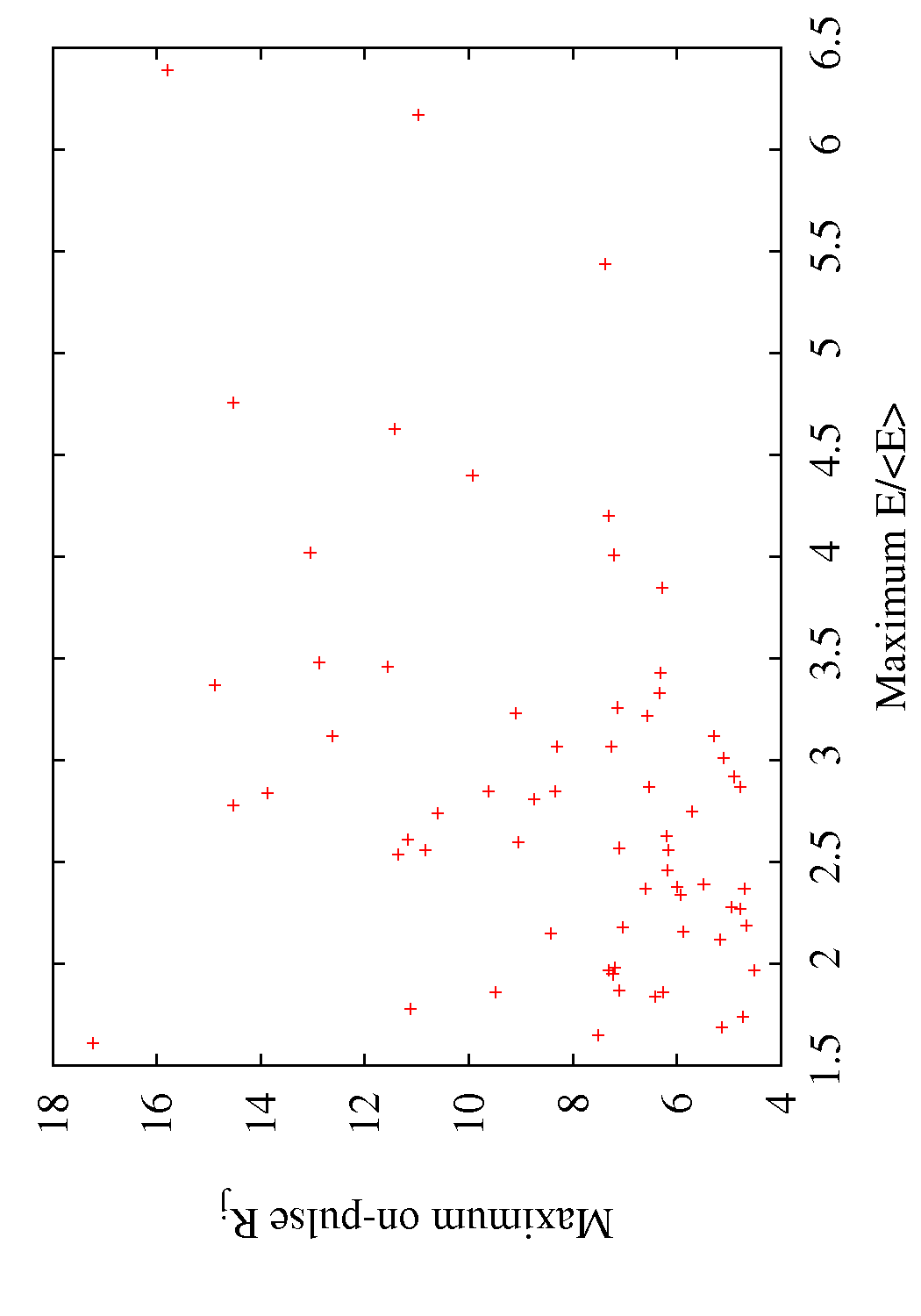}
\caption{The $R$-parameter plotted against the maximum deviation of integrated normalised pulse energy.}\label{fig:remax}
\end{figure}

\subsection{Energy budgets and additional insight from interpulse pulsars}\label{sec:interps}
We find it striking that for $\se>4$ pulsars, the maximum deviation of integrated pulse energy from $\enorm$ tends to be fairly low. Inspecting the maximum integrated energy deviation ($M_{\rm E}$) in these pulsars, we find that they range $1.6<M_{\rm E}<6.4$, with a mean of 2.9; that is, the integrated pulse energy tends to not deviate vastly from its mean value. 
One might expect maximum $\rj$ to correlate with $M_{\rm E}$, given the excess energy one would expect to be provided by a bright, phase-resolved pulse. In Figure \ref{fig:remax}, we show a scatter plot of maximum $\rj$ vs. $M_{\rm E}$ for pulsars with $\se>4$ and a significantly measured $\rj$ value. While there is a weak correlation here (Kendall's tau test gives $\tau=0.29, P_\tau=0.001$), the scatter in both variables is significant.

This scatter and the fact that $\rj$ for some pulsars is large across a broad phase range indicates that many phases may be emitting large bursts of energy; as previously noted, phase-resolved changes in $\rj$ indicate the possibility that the energy distribution is likewise phase-dependent. Despite this, however, we see only a small range of $M_{\rm E}$ values, and at least 45\% of our sample has an integrated pulse energy distribution that is well-fit to a unimodal (mostly log-normal) distribution. What this appears to imply is that despite the occurrence of sizeable sub-beam variations, a large outburst at one phase is compensated by a deficit of or weakened emission at other phases, such that a narrow integrated distribution in energy may be maintained. Thus, there appears to be a self-balancing effect, i.\,e. there is a net energy regulation by which the total sub-beam circuit is governed. 
% that the different sub-beams might know about some kind of pulsar energy budget?
%One might expect $R$ to correlate with $M_E$, given the excess energy one would expect to be provided by a large pulse. In Figure \ref{fig:remax}, we show a scatter plot of $R$ vs. $M_E$ for pulsars with $\se>4$ and a significantly measured $R$ value. 

Similarly, previous studies have indicated that pulsars with interpulses show a relationship in the pulses' emission properties. Various studies have shown correlations or anti-correlations in the amplitude of main pulses and interpulses \citep[e.\,g.][]{fowlerwright82,biggs90}. Furthermore, \citet{patricketal06} found the same periodicity of amplitude modulation in the main/interpulse of PSR\,J1705--1906.
%This concept can be further explored by considering the pulsars with an interpulse, which represents a discrete emission region rather than a sub-beam of the main pulse. \citep{patricketal06} noted a number of  

We identified five interpulse pulsars in our sample; the main and interpulses for these pulsars (``interpulse'' here being the fainter component) have separately reported statistics in Table \ref{table:stats}, marked by (m) and (i), respectively, in addition to the statistics from the total integrated emission window. We find agreement between maximum $\rj$ in the main pulse and interpulse only in the case of PSR\,J1705--1906, which despite a factor of $\sim$7 difference in emission intensity, the maximum $\rj$ values both peak from 8$-$10. This is particularly notable as it supports the aforementioned findings of \citet{patricketal06} and \citet{patricketal07}.

In the other interpulse pulsars, all but PSR\,J0908--4913 exhibit $\rj$ values significant only in the main pulse. We note that even in the presence of a pulsar-wide energy regulation, these discrepancies may not be surprising given the strong phase-dependence of $\rj$ and thus its implied dependence on viewing angle. Accordingly, it may be that we view PSR\,J1705--1906's main pulse and interpulse at an angle such that we see corresponding primary and counter-beam components; while the other pulsars might share properties between particular sub-beams, their properties could be masked by an unfavourable viewing angle.

The energy distribution classification differences between main pulse, interpulse, and net emission is also interesting to consider in this discussion. However, due to the low $\se$ on all of the interpulses, the data do not provide clear results on this topic. Most of the classifications are ``other'', and only the interpulses of PSR\,J1705--1906, J1739--2903, and J1825--0935 are well-fit to a log-normal distribution. While this seems to imply that the main pulse and interpulse energy distributions do not share the same underlying plasma statistics, higher signal-to-noise measurements would be more suitable to explore the relationship between the main/interpulse integrated energy distribution. 

\subsection{Notes on anomalous pulsar properties}\label{sec:weirdos}
In our online figure (see Appendix A), we display the pulse stack, modulation and intensity profiles, and energy distribution with fits for each pulsar. These graphics provide a wealth of information, and upon viewing them, nearly every pulsar appears to have some unique and fascinating feature. As such, the plots contain far more features of interest than are relevant for the discussion in this paper. The reader is encouraged to inspect the data and pursue outstanding features that catch their interest. Examples of peculiar behaviour which stood out to us are: the cyclic, phase-dependent nulling of PSR\,J1133--6250, the atypically broad energy distributions of PSRs J1243--6423, J1047--6709, J1401--6357, J1456--6843, and J1745--3040, and the ordered beating visible in the pulse stack of PSR\,J1534--5334). 
%I'll mention the drifters, since those are probably more generally of interest and at least represent a well-defined oddity. Anyways, I think I will end that section with some sentence like "the reader is encouraged to inspect this data and pursue outstanding features that catch their interest".

Below, we do describe unconventional emission discovered in several pulsars, for cases where the anomalous behaviour is not recognisable from the displayed data.

\subsubsection{Multi-state nulling fraction pulsars}
\citet{sbsmb} reported what appeared to be a ``part-time RRAT'', PSR\,J0941--39, which at times is observable as a nulling pulsar with a null fraction of $\sim$10\%, and at other times emits single pulses at a rate of $\sim$2 per minute. Our analysis has identified several potential further examples of these. The first is PSR\,J0828--3417, which was originally noted to have low-level emission during its ``nulls'' by \citet{esamdinetal05}. We note that in fact their ``low-level emission'' appears to be made up of sporadic single pulses, similar to PSR\,J0941--39. We furthermore discovered one pulse from PSR\,J1107--5907, which has been previously reported as an ``intermittent pulsar'' whose emission alternates between states of bright, weak, and null emission on a yet-unknown timescale \citep{kramerintermittentreview,obrienthesis}. Archival data from this pulsar also reveals erratic changes in nulling fraction, particularly directly preceding its constant on state. We think it pertinent to point out that each of these pulsars exhibit changes not only in intensity (i\,e. nulls or mode changes when emission appears to cease), but also in the \emph{relative time spent} in on- and off-configurations.
% Fuller discussion on this and any figures should be saved for its own paper, I think, or discussion in a later full paper about PSR\,J0941--39.

\subsubsection{Wide nulling distribution of PSR\,J1255--6131}
We detected only one pulse from PSR\,J1255--6131, and no integrated emission. The single pulse was of high significance (${\rm S/N}\sim10$), and thus given the apparently large nulling fraction, we were motivated to explore archival data for this pulsar. We found $\sim$20 archival pulse stacks at a central frequency of 1.4\,GHz. These observations were collected as follow-up to the Parkes Multibeam survey, and the data format and observing system is as described in the survey's paper \citep{pksmb}. The data available to us were each of length 5 to 30\,minutes, formed into pulse stacks and often averaged over 1-minute intervals so we could not probe single-pulse behaviour. We found that PSR\,J1255-6131 displayed a great range of activity cycle times in these observations. Occasionally the pulsar appeared to emit without ceasing over lengths of 5 to 10 minutes, however more commonly it exhibited bursts of emission lasting up to 3\,minutes. The null fraction from observation to observation ranged between 10 and 100\%. We estimate the null fraction for the pulsar is typically around $\sim$70-80\%, however with a broad emission cycle range.
%Tob Tnav  Tfav  NF
%20	3	4	55
%20	<1	14	85
%20	<1	4	75
%10	2	8	80
%10	4	6	60
%10	<1	9	90
%10	5	5	50
%8	8	0	0
%6	0	20	100
%6	0.5	5.5	92
%6 off
%6 off
%30	5	?	70%
This pulsar may be exhibiting a variation of the null-change behaviour discussed in the above section. However, the behaviour in this pulsar differs as its nulling fraction distribution does not appear to be bimodal, but rather is an unusually broad, and may be either stochastic, or a smoothly-distributed function. The HTRU med-lat survey pointing appears to have caught PSR\,J1255--6131 at the sparsest tail of this null fraction distribution.

%(J1732-4156?)
\subsubsection{Off-pulse emission and PSR\,J1406--5806}
The careful cleaning of interference in our data enabled us sensitivity to short-duration, off-pulse emission (i\,e. emission more than 5\% in phase from integrated profile components). We found such emission in only one pulsar, PSR\,J1406--5806. Pulses were detected across the full phase range, though its ``on-pulse'' emission appears also to be highly sporadic. Upon stacking all available archival pulse stacks from Parkes Telescope (3\,h total), the off-pulse emission contributes weak components to a stable integrated profile. We interpret this object as an aligned rotator with a high nulling fraction. The lack of off-pulse emission in other pulsars, indicates no evidence that the emission detected by \citet{offpulseemission} is made up of sporadic, bursty emission.

\section{Summary and Conclusions}\label{sec:conclusions}
We analysed the pulse-to-pulse energy and modulation properties of all pulsars serendipitously observed in the HTRU intermediate latitude survey to emit detectable single pulses. This sample was derived from the 702 pulsars re-detected by the HTRU med-lat observations, yielding this survey a single-pulse detection rate of 45\%. 16 of these pulsars were only detected through the single-pulse search. 

For our full 315-pulsar sample, we performed energy distribution fits to determine the suitability of log-normal or Gaussian distributions to describe integrated pulse-to-pulse energy. This analysis showed that more than 40\% of our sample fits a log-normal pulse energy distribution, while only a few pulsars were well-fit to a Gaussian distribution. Other pulsars were not fit by either distribution, however this may be due to $a$) the influence of noise on a faint pulsar; $b$) unrecognised nulling/mode-changing; or $c$) a non-Gaussian, non-log-normal underlying energy distribution. Because of the large likelihood of $a$ and $b$ to disrupt our fits, we suggest that a greater fraction of pulsars may show unimodal and log-normal pulse energy distributions, however observations of higher sensitivity and a greater number of detected pulses will be required to address this.

Some pulsars were found to have bi- or multi-modal energy distributions, which we demonstrated to be caused by mode changes in some pulsars. Energy distribution inspection can thus be useful for identifying mode changes where they might not be obvious in a pulse stack. Multi-energy states have implications for nulling pulsars, supporting the argument of \citet{nullsaremodes} that some pulsars observed to ``null'' may simply be reconfigured into a state where fainter, fewer, or no sub-beams are directed at the observer. Along these lines, we demonstrated that the previously ``nulling pulsar'' PSR\,J1900-2600 exhibits faint emission in its low state. It should be noted that for some multi-modal energy pulsars, particularly those with short change timescales or those with minimal differences in mode energies, we cannot distinguish between mode changing and other longitude-resolved modulation (e.\,g. a distinct transient sub-beam).

Mode-changing properties are not to be confused with another state-change effect observed in only a few pulsars. PSRs J0828--3417 and J1107--5907 appear to have two discrete nulling fraction states; as with PSR\,J0941--39 \citep{sbsmb}, these objects switch between being pulsars with null fraction $<$10\%, to a separate state where they sporadically emit single pulses per many rotations. The single-pulse state may have previously been falsely identified as a low-energy mode change \citep{esamdinetal05,obrienthesis}, as during long integrations bright single pulses are dampened by the addition of null rotations.

We reported longitude-resolved modulation statistics, quantified by the modulation index $m$ and the $R$-parameter, the latter used to identify non-Gaussian sporadic emission. We found no correlations between $m$ or $R$ and physical parameters (e.\,g. age, spin-down energy) or the complexity parameters predicted by various energy models.
% However, existing correlations may be diluted by the mixing of conal and core-type profiles.

We found that outbursts with distribution significantly deviating from Gaussianity \emph{can} occur across a pulsar's full integrated profile, however occasionally this modulation intensifies at the rising and/or falling edge of components in the integrated profile. This may indicate a conal or core edge modulation effect, and/or the presence of giant pulses at the corresponding pulse phase. Our analysis supports the fact that physically distinctive ``giant pulse'' phenomena should be defined thus because they have power-law statistics, however we have insufficient data to assess the shape of high-energy tails in our high-$R$-parameter pulsars.

Finally, in considering the large phase-resolved deviations seen in the $\rj$ profiles of some pulsars in conjunction with the distribution of integrated single-pulse energy in these pulsars, it is striking that the energy typically deviates only up to three times its average value. This is suggestive of a beam-wide energy regulation that affects all angles of the pulsar's beam, and we broaden our consideration of this possibility using information from interpulse statistics. Only one of our five interpulse pulsars exhibits a notable relationship in modulation properties between the main and interpulse, however we cannot make conclusive statements about the main/interpulse energy distribution relationship due to the low signal from the interpulses in these pulsars.

\section{Ackowledgements}
%SBS acknowledges that she likes corned beef hash.
%SBS acknowledges the support of NASA's Jet Propulsion Laboratory to  
The Parkes Radio Telescope is part of the Australia Telescope National Facility which is funded by the Commonwealth of Australia for operation as a National Facility managed by CSIRO. A portion of research was carried out at the Jet Propulsion Laboratory, California Institute of Technology, under contract with the National Aeronautics and Space Administration.

\bibliographystyle{mn2e}
\bibliography{knownHTRU}

\appendix
\section{Online Supplementary Material}\label{sec:appendix}
Here we describe the supplementary material available linked to this paper on the MNRAS website.
Before full publication in MNRAS the material may be downloaded at \mbox{http://dl.dropbox.com/u/22076931/supplementary\_material.pdf}

\begin{figure*}
\centering
\includegraphics[width=0.99\textwidth,trim=0mm 0cm 0cm 0cm, clip]{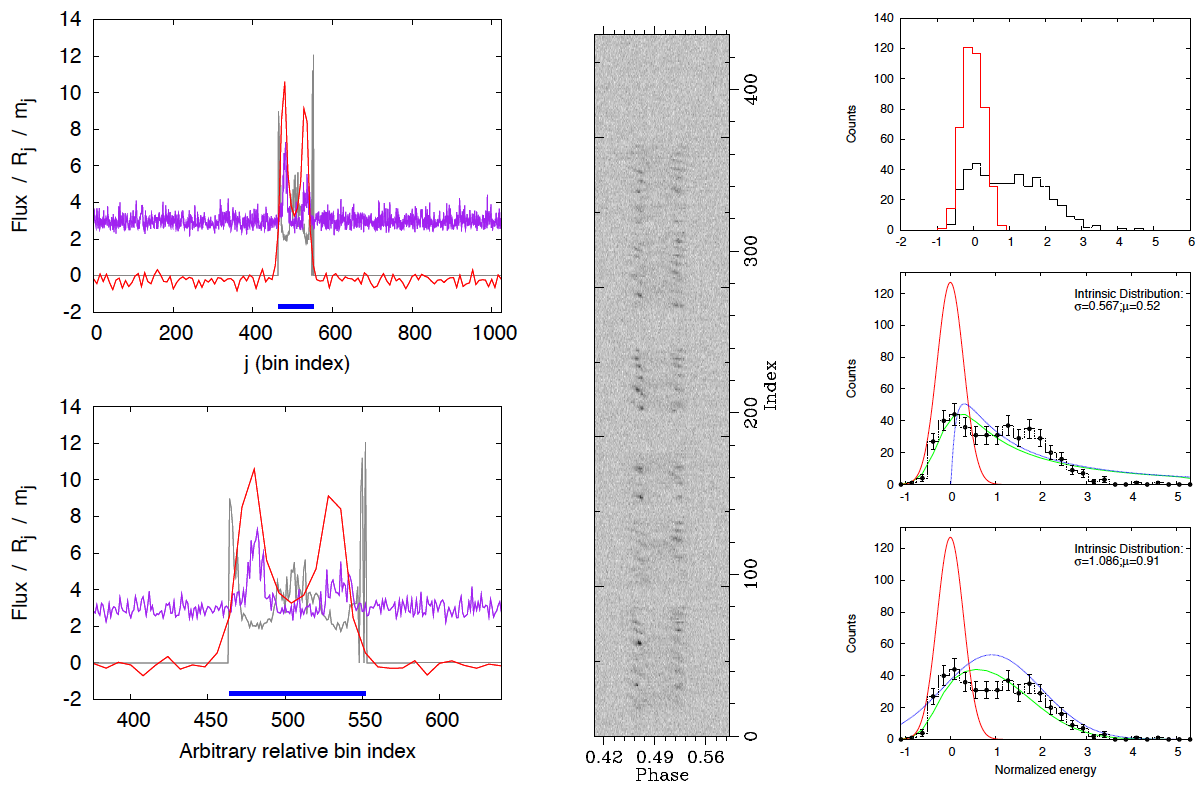}
\caption{Here we show one example of a plot from the online supplementary material. These panels show the data corresponding to pulsar PSR\,J1727-2739. In addition to the figures shown here, the supplementary material also provides the pulsar's J2000 name, dispersion measure and period, and the observation's start time in Universal Coordinated Time for reference. See text for panel descriptions.}\label{fig:supfig}
\end{figure*}

The online material provides detailed data viewgraphs for each of the 315 pulsars in our sample, as shown in Figure \ref{fig:supfig}. For each pulsar, the figure panels are as follows: \emph{Upper left}: The total intensity profile integrated over the full observation and integrated in $j$ to maximise the profile S/N (red), the longitude-resolved $R$-parameter, $R_j$ (purple), and the longitude-resolved modulation index, $m_j$ (grey), where $0\leq m_j\leq 13$ (this value is generally otherwise undefined or badly constrained). The latitude range defined as ``on-pulse'' for the purposes of our analysis is marked by one or more horizontal blue bars. \emph{Lower left}: as with the upper left plot, however zoomed in to accent features in the on-pulse region. \emph{Centre}: The pulse stack, where the observed power is represented in greyscale as a function of pulse phase and number (indexed from the observation start). \emph{Upper right}: The observed on- and off-pulse energy distributions (in black dash-dot and red solid lines, respectively). \emph{Centre and lower right}: the energy distribution of on-pulse data (black dash-dot histogram with points and error bars), the noise-convolved model fits (thick green line), the off-pulse noise model (thin red line), and the intrinsic energy distribution (blue dotted line) for the best-fit log-normal and Gaussian distributions (centre and right, respectively). Errors shown for the on-pulse data are the square root of the number in each bin.

\end{document}